\documentclass[pre,aps,eqsecnum,preprint]{revtex4}
\usepackage{graphicx}
\usepackage{amsmath}
\usepackage[cal=Boondox]{mathalfa}
\usepackage{slashed}
\usepackage[dvips]{epsfig}
\usepackage{color}
\definecolor{red}{rgb}{1,0,0}
\definecolor{green}{rgb}{0.13,0.55,0.13}
\definecolor{blue}{rgb}{0,0,1}

\newcommand{\red}[1]{{\color{red} #1}}
\newcommand{\blue}[1]{{\color{blue} #1}}

\begin{document}
	
\title{\large Interplay of order and disorder in two-dimensional critical systems with mixed boundary conditions}
	
	\author{E. Eisenriegler}
	
	\affiliation{Institute for Advanced Simulation, Theoretische Physik der lebenden Materie (IAS-2), Forschungszentrum J\"ulich, D-52425 J\"ulich, Germany}

	\date{\today}

	\begin{abstract}
In spin systems such as the Ising model, the local order and disorder can be characterized by the order-parameter and energy density profiles $\langle \sigma ({\bf r}_1) \rangle$ and $\langle \epsilon ({\bf r}_2) \rangle$, respectively. Does increasing the order at ${\bf r}_1$ always decrease the disorder at ${\bf r}_2$? Does increasing the disorder at ${\bf r}_2$ always decrease the order at ${\bf r}_1$? The answer to these questions is contained in the cumulant response function $\langle\sigma ({\bf r}_1) \, \epsilon ({\bf r}_2) \rangle^{(\rm cum)}$. This correlation function vanishes in the unbounded bulk but not in systems with fixed-spin boundary conditions. Using the universal operator-product expansion of $\sigma ({\bf r}_1) \, \epsilon ({\bf r}_2)$ and exact results for the Ising model, we analyze $\langle\sigma ({\bf r}_1) \, \epsilon ({\bf r}_2) \rangle^{(\rm cum)}$ in two-dimensional critical systems defined on the $x-y$ plane with mixed $+$ and $-$ boundary conditions. Particularly interesting behavior is found when either of the operators $\sigma$ or $\epsilon$ is located on a ``zero line" in the $x-y$ plane, along which $\langle\sigma ({\bf r})\rangle$ vanishes. Results for half-plane, triangular, and rectangular geometries are presented.

\vspace{2cm}

Published in Physical  Review E {\bf 113}, 024105 (2026)

	\end{abstract}
	
	\maketitle

\section{INTRODUCTION} \label{intro}
Order and disorder are basic concepts in the field of phase transitions and critical phenomena.  Interesting effects arise in critical systems with boundaries \cite{Diehl,CardBoundaryCritPhen}. Here we investigate the interplay of order and disorder in the two-dimensional Ising model with mixed boundary conditions \cite{BX,BG,BE21,E23} right at its bulk-critical point. In Ising systems the local order and disorder can be characterized by the averages of the density operators $\sigma ({\bf r})$ and $\epsilon({\bf r})$ of the order parameter and the energy which are odd and even, respectively, under the change of direction of all Ising spins. In the conformal classification \cite{BPZ}, $\sigma ({\bf r})$ and $\epsilon({\bf r})$ are the only primary operators of the two-dimensional Ising model. 

For a given configuration of Ising spins, $\sigma ({\bf r})$ is, in the sense of coarse graining \cite{ferro}, proportional to the difference of the number of up spins and down spins in a volume element around ${\bf r}$ while $- \epsilon ({\bf r})$ is proportional to the sum of the products of nearest neighbor spins in the volume element with its bulk-average subtracted \cite{endens}. Thus, $\langle \sigma ({\bf r}) \rangle$ specifies the direction and magnitude of the local spin alignment, and provides a measure of the local order. The quantity $\langle \epsilon ({\bf r}) \rangle$ vanishes in the bulk and increases as the correlation between neighboring spins decreases and thus characterizes the local disorder \cite{shortraor}. 

It is convenient to normalize the two operators via their two-point function in the bulk or, equivalently, via the leading term in their operator product expansion ``OPE'' \cite{BPZ,BE21} such that
\begin{eqnarray} \label{selfOPE} 
\sigma({\bf r}_1)  \sigma({\bf r}_2) \to {1 \over |{\bf r}_{12}|^{1/4} } \, + \, ... \, , \quad \epsilon({\bf r}_1)  \epsilon({\bf r}_2) \to {1 \over |{\bf r}_{12}|^2 } \, + \, ... \, .
\end{eqnarray}
Here ${\bf r}_{12} \equiv {\bf r}_1 - {\bf r}_2 \equiv - {\bf r}_{21}$.

Our goal is to investigate the response $\delta \langle \epsilon ({\bf r}_2) \rangle$ of the disorder $\langle \epsilon \rangle$ at point ${\bf r}_2$ to a weak up ordering imposed at point ${\bf r}_1$. This is given by the response function
\begin{eqnarray} \label{respfunction}
\langle \sigma ({\bf r}_1) \epsilon ({\bf r}_2) \rangle^{(\rm cum)} \equiv \langle \sigma ({\bf r}_1) \epsilon ({\bf r}_2) \rangle - \langle \sigma ({\bf r}_1) \rangle \langle \epsilon ({\bf r}_2 \rangle \, .
\end{eqnarray}
This same function also describes the response $\delta \langle \sigma ({\bf r}_1) \rangle$ of $\langle \sigma \rangle$ at point ${\bf r}_1$ to a weak increase of disorder imposed at point ${\bf r}_2$.  

The response function (\ref{respfunction}) vanishes in the unbounded bulk, due to the $\sigma \to -\sigma$ symmetry. In systems with boundary spins fixed in the up or down direction, this symmetry is broken and the response function is nonvanishing.

Before analyzing the response function (\ref{respfunction}) in detail, we consider some {\it intuitive} expectations. As for the geometry, a useful example is the upper half plane bounded by the $x$ axis.

(i) For a uniform + boundary, $\langle \sigma\rangle$ is positive everywhere in the upper half plane. Increasing the up-ordering, e.g. by applying a local magnetic field at ${\bf r}_1$, also tends to further align spins in the surroundings in the up direction. Thus, both $\delta \langle \epsilon ({\bf r}_2) \rangle$ and the response function (\ref{respfunction}) are negative. The same conclusion follows from increasing the disorder at ${\bf r}_2$, e.g. by local heating, which reduces the up alignment  there and in the surroundings, so that $\delta \langle \sigma ({\bf r}_1) \rangle$  and the response function (\ref{respfunction}) are negative. With the same type of arguments or from simply reversing the directions of all the spins, one finds that for  a uniform $-$ boundary the response function  (\ref{respfunction}) is positive . For either sign of the boundary spins, increasing the magnitude of the order at ${\bf r}_1$ decreases the disorder at ${\bf r}_2$, and increasing the disorder at ${\bf r}_2$ decreases the magnitude of the order at ${\bf r}_1$.

(ii) What happens for mixed $+-$ boundary conditions \cite{BX}, where the boundary spins are fixed in the up and down direction along the negative and positive $x$ axis, respectively? Here, the $y$ axis separates regions of positive and negative order and represents a ``zero line'', along which $\langle \sigma \rangle$ vanishes \cite{BX,BG}. A local up-ordering imposed at a point ${\bf r}_1$ right on the zero line extends into the surroundings on the left and right, generating, by the arguments given in (i), a decrease and increase, respectively, of the disorder at ${\bf r}_2$. This leads to an interesting behavior of the response function (\ref{respfunction}), discussed below.

(iii) Placing the point ${\bf r}_2$ of disordering right on the zero line leads to another interesting effect. One might guess that the increase in disorder on the zero line extends to both sides of the line, thereby {\it reducing} the magnitude of the order on both sides, but this is incorrect. The magnitude of the order at ${\bf r}_1$ is actually {\it increased}, as we shall see. 

Predictions (i)-(iii) can be put on a sound footing without detailed calculations by invoking the OPE \cite{higher} of $\sigma \times \epsilon$. For this, only the first two terms of the expansion are needed which read
\begin{eqnarray} \label{next}
\sigma ({\bf r}_1) \epsilon ({\bf r}_2) &\to& - {1 \over 2 |{\bf r}_{12}|} \Bigl[1+4 \, {\bf r}_{21} \, \partial_{{\bf r}_1} + ...  \Bigr] \sigma ({\bf r}_1) \nonumber \\
&\to& - {1 \over 2 |{\bf r}_{12}|} \Bigl[1 - 3 \, {\bf r}_{12} \, \partial_{{\bf r}_2} + ...  \Bigr] \sigma ({\bf r}_2)  \, .
\end{eqnarray}
The two displayed expressions on the right hand side of (\ref{next}) are equal, apart from higher order terms in the small distance $|{\bf r}_{12}|$. This follows from $\sigma({\bf r}_1) \to [1+{\bf r}_{12} \, \partial_{{\bf r}_2}]\sigma ({\bf r}_2)$. There is an important difference between the OPE in (\ref{next}) and those in (\ref{selfOPE}). In Eq. (\ref{selfOPE}) the leading contributions are given by the non-fluctuating terms $|z_{12}|^{-1/4}$ and $|z_{12}|^{-2}$, which are the same as in the bulk spin-spin and energy-energy correlation functions and which are unperturbed by the boundary conditions or other operators. In Eq.~(\ref{selfOPE}) these perturbations enter in next-to-leading order, but in the OPE (\ref{next}) they enter in {\it leading} order \cite{spinreverse}.  

Taking the thermal average of Eq. (\ref{next}), one obtains exact, universal expressions for the short-distance behavior \cite{singularope} of the response function (\ref{respfunction}) that confirm and extend the above intuitive expectations.

Case (i): For original order $\langle \sigma ({\bf r}_1) \rangle >0$ and $\langle \sigma ({\bf r}_1) \rangle <0$, the first terms on the right-hand side of (\ref{next}) predict $\langle \sigma  \epsilon \rangle^{(\rm cum)} <0$  and $\langle \sigma  \epsilon \rangle^{(\rm cum)} >0$, respectively, in agreement with the intuitive arguments in (i). We also obtain an explicit expression for the power-law singularity in the short-distance behavior.

Case (ii)  If $\langle \sigma ({\bf r}_1) \rangle$ vanishes, the leading behavior for small $|{\bf r}_{12}|$ is determined by the second term in (\ref{next}), provided that the first derivative $\partial_{{\bf r}_1} \langle \sigma ({\bf r}_1) \rangle$ is nonvanishing. In the $+-$ half plane let us place $\sigma$ right on the zero line, choosing ${\bf r}_1 = (x_1,y_1)=(0, y_0)$ \cite{mixup},  and let us place $\epsilon$ on the line $y=y_0$. Then, the upper Eq. (\ref{next}) yields
	\begin{eqnarray} \label{x1zero}
\langle \sigma (0 , y_0) \, \epsilon (x_2 , y_0 ) \rangle_{+-}^{(\rm cum)} \to 2 \, \bigl({\rm sign}\,x_2 \bigr) \times  \big|\bigl[ \partial_{x_1} \langle \sigma (x_1 , y_0) \rangle_{+-} \bigr]_{x_1=0} \big| 
	\end{eqnarray}
for small $|x_2|$. The signs of $\langle \sigma \, \epsilon \rangle_{+-}^{(\rm cum)}$ in (\ref{x1zero}) are in agreement with the decrease and increase of disorder to the left and right of the zero line, as argued intuitively in (ii). Instead of the power law divergence in (i), the nonanalyticity for $x_2 = x_1$ now has the milder form of an upward jump in sign as $x_2$ grows. Later on we shall consider cases where both $\langle \sigma  \rangle$ and its first derivative vanish, leading to an even milder form of nonanalyticity. 

Case (iii): For $\epsilon$ right on the $y$-axis, i.e., ${\bf r}_2=(x_2,y_2)=(0,y_0)$, the lower form of Eq. (\ref{next}) yields a {\it downward} jump in sign
\begin{eqnarray} \label{x2zero}
\langle \sigma (x_1 , y_0) \epsilon (0, y_0 ) \rangle_{+-}^{(\rm cum)} \to - (3/2) \, \bigl({\rm sign}\,x_1 \bigr) \times  \big|\bigl[ \partial_{x_2} \langle \sigma (x_2 , 0) \rangle_{+-} \bigr]_{x_2=0} \big|
\end{eqnarray}
as the position $x_1$ of $\sigma$ increases along the line $y=y_0$. Thus, increasing the disorder right on the defect line  {\it enhances} the up-order to its left and the down-order to its right. The reason is that near the $y$-axis, where the sign of $\langle \sigma \rangle_{+-}$ changes, the disorder is large. Further increasing the disorder facilitates the sign change and leads to the enhancement. This argument is made more explicit in Ref. \cite{explain}. 

While the upward and downward jumps depend on the value of the first derivative $\partial \langle \sigma \rangle$ this drops out from their ratio $2/(-3/2) = -4/3$ which is universal, i.e., the same for crossing other zero lines of $\langle \sigma \rangle$. Examples are presented in sections
\ref{hp-+-} and \ref{equitri}.

So far we have only analyzed the response function $\langle\sigma({\bf r}_1)  \epsilon({\bf r}_2)\rangle$ for short distances between ${\bf r}_1$ and ${\bf r}_2$. In Sections \ref{uniform} -\ref{equitri} we study the response function for arbitrary ${\bf r}_1$ and ${\bf r}_2$ in some Ising systems with uniform and mixed boundary conditions for which exact results \cite{BX,BG} are available. These include the upper half planes with  $+$, $+-$, and $-+-$ boundary conditions and a finite equilateral triangle with $-+-$ boundaries. This gives us the opportunity to confirm that the response function not only obeys the OPE in Eq.~(\ref{next}) and its extension to higher order in Appendix \ref{OPE3} but also the 
`boundary-operator expansion'' BOE \cite{Diehl,CardBoundaryCritPhen,BE21} and the `corner-operator expansion'' COE \cite{E23}.

Finally, in Section \ref{-+-+square} we consider a square with $+$ spins on the horizontal boundaries and $-$ spins on the verticle boundaries, at the center of which both $\langle \sigma \rangle$ and its first derivatives vanish. Here the OPE predicts that the leading short-distance singularities of the response function have a cusp-like form.

\section{Upper half plane with a uniform boundary condition +} \label{uniform}

In the upper half plane with uniform boundary condition + the two point function reads 
\begin{eqnarray} \label{+epssig}
&& \qquad \qquad \langle \sigma (1) \epsilon (2) \rangle_+ = \langle \sigma (1) \rangle_+ \, \langle \epsilon (2) \rangle_+ \times {1 \over |{\bf r}_{12}|} \, {(x_1 - x_2 )^2 + y_1^2 + y_2^2 \over \sqrt{(x_1 - x_2 )^2 + (y_1 + y_2)^2} } \, \nonumber \\
&&\langle \sigma (1) \rangle_+ = (2/y_1)^{1/8} \, , \, \langle \epsilon (2) \rangle_+ =-1/(2y_2) \, , \, |{\bf r}_{12}|=\sqrt{(x_1 - x_2 )^2 + (y_1 - y_2)^2} \, ,
\end{eqnarray}
see Ref. \cite{BX}. For short we here denote ${\bf r}_1 =(x_1, y_1)$ and ${\bf r}_2 = (x_2, y_2)$ by 1 and 2, respectively. For $|{\bf r}_{12}| \to 0$ the two-point function approaches 
\begin{eqnarray} \label{+epssigclose}
\langle \sigma (1) \epsilon (2) \rangle_+ \to - \langle \sigma (1) \rangle_+  \times {1 \over 2 |{\bf r}_{12}|} \, ,
\end{eqnarray}
consistent with the OPE (\ref{next}), and for $|{\bf r}_{12}| \to \infty$ with $y_1 , \, y_2$ fixed it approaches $\langle \sigma (1) \rangle_+ \langle \epsilon (2) \rangle_+ [1+2(y_1 y_2 /|{\bf r}_{12}|^2)^2]$, consistent with the BOE \cite{BOE} together with the form $\langle T(x_1) T(x_2) \rangle_+ = 1/[4 (x_1 -x_2)^4]$ of the $TT$ cumulant in the boundary. Here $T(z)$ is the stress tensor \cite{CardBoundaryCritPhen}.

For later comparison we note the corresponding cumulant along the horizontal axis $y=y_0$
\begin{eqnarray} \label{+epssigprime}
\langle \sigma (x_1, y_0) \epsilon (x_2, y_0) \rangle_+^{\rm (cum)} = -  \, {2^{(1/8)-1} \over y_0^{(1/8)+1}} \times \Bigl[|X_{21}|^{-1} (X_{21}^2 +4)^{-1/2}  (X_{21}^2 +2) \, - \, 1\Bigr] \, 
\end{eqnarray}
where $X_{21}=(x_2 -x_1)/y_0$. It is an even function of $X_{21}$ that, starting from $- \infty$, monotonically increases to 0 when $|X_{21}|$ increases from 0 to $\infty$. Correspondingly the odd  second term in the OPE (\ref{next}) is absent since $\partial_{x_1} \langle \sigma (1) \rangle_+$ vanishes. 

This is different along the vertical axis $x=0$ where 
\begin{eqnarray} \label{+epssigprimeprime}
\langle \sigma (x_1 =0, y_1) \epsilon (x_2 =0, y_2) \rangle_+ = \langle \sigma (1) \rangle_+ \, \langle \epsilon (2) \rangle_+ \times {y_1^2 +y_2^2 \over |y_1^2 - y_2^2|} \, .
\end{eqnarray}
Here both even and odd powers of $y_{21}$ appear and the nonvanishing second term in (\ref{next}) is reproduced. It is rewarding to check the consistency between (\ref{+epssigprimeprime}) and the OPE to higher order. This is done up to order $y_{21}^3 / |y_{21}|$ in Appendix \ref{opuniform}.

\section{Upper half plane with a $+-$ boundary condition} \label{hp+-}

The upper half $z=x+i y$ plane with boundary condition + for $x < 0$ and $-$ for $x>0$ is perhaps the simplest geometry that displays the features mentioned in paragraphs (ii) and (iii) in the Introduction. For the present $+-$ boundary condition we can use the explicit form of the response function derived by Burkhardt and Xue, see Eqs. (4.1) and (4.3) in Ref. \cite{BX}. First note its basic symmetry
\begin{eqnarray} \label{+-symmetry} 
\langle \sigma (x_1, y_1) \, \epsilon (x_2, y_2) \rangle_{+-}^{(\rm cum)} =  - \langle \sigma (-x_1, y_1) \, \epsilon (-x_2, y_2) \rangle_{+-}^{(\rm cum)}
\end{eqnarray}
which implies, in particular, that it vanishes for $x_1 =x_2 =0$. Now consider the response function along the line $y=y_0$ with one of the two operators on the zero line $x=0$, i.e. $x_1 =0$, $x_2$ arbitrary (case (i)) and $x_2 =0$, $x_1$ arbitrary (case (ii)).
	
In case (i), $\langle \sigma \rangle_{+-}$ vanishes and one finds
\begin{eqnarray} \label{half+-} 
\langle \sigma (0, y_0) \, \epsilon (x_2, y_0) \rangle_{+-}^{(\rm cum)} =  {2^{(1/8)-1} \over y_0^{(1/8)+1}} \times 4 \, \Biggl(1+{X_2^2 \over 4}\Biggr)^{-1/2} \, (1+X_2^2)^{-1} \,  {\rm sign} X_2 \, ,
\end{eqnarray}
where $X_2 \equiv x_2 / y_0$ and, for comparison, we have used the same prefactor as in (\ref{+epssigprime}). In accordance with (\ref{+-symmetry}), Eq. (\ref{half+-}) is odd in $X_2$ and it shows the limiting behaviors
\begin{eqnarray} \label{half+-limits} 
\langle \sigma (0, y_0) \, \epsilon (x_2, y_0) \rangle_{+-}^{(\rm cum)} \to {2^{(1/8)-1} \over y_0^{(1/8)+1}} \times 4 \, \Bigl\{({\rm sign} X_2) \Bigl( 1- {9 \over 8} X_2^2 \Bigr)  \, , \, {2 \over  X_2^3}  \Bigr\} \, 
\end{eqnarray}
for $|X_2| \to \{ 0 \, , \, \infty \}$. These are in agreement with the predictions of the $\{$OPE , BOE$\}$. For the OPE prediction in leading order see Eq. (\ref{x1zero}) together with (\ref{sig+-}) and in next-to-leading order see Eq. (\ref{nextto+-}). For the BOE prediction see (\ref{Tsigma}) together with Ref. \cite{BOE}. 

In case (ii) the result is
\begin{eqnarray} \label{half+-prime} 
&&\langle \sigma (x_1, y_0) \, \epsilon (0, y_0) \rangle_{+-}^{(\rm cum)} = - {2^{(1/8)-1} \over y_0^{(1/8)+1}} \times 3 \, \Bigl(1+{X_1^2 \over 2 }  \Bigr) \Biggl(1+{X_1^2 \over 4}\Biggr)^{-1/2} \times \nonumber \\
&& \qquad \qquad \times (1+X_1^2)^{-1/2} \,  {\rm sign} X_1 \, - \, \langle \sigma (x_1, y_0) \rangle_{+-} \times \langle \epsilon (0, y_0) \rangle_{+-} \, ,
\end{eqnarray}
where
\begin{eqnarray} \label{half+-onepoint} 
\langle \sigma (x_1, y_0) \rangle_{+-} = - \Bigl( {2 \over y_0} \Bigr)^{1/8} \, (1+X_1^2)^{-1/2} \, X_1 \, , \quad \langle \epsilon (0, y_0) \rangle_{+-}= {3 \over 2 y_0} \, ,
\end{eqnarray}
and where $X_1 =x_1 / y_0$. The two contributions to the cumulant, the two-point function $\langle \sigma \epsilon \rangle_{+-}$ and the subtracted product of one-point functions $-\langle \sigma \rangle_{+-} \times \langle \epsilon \rangle_{+-}$, are both odd in $X_1$ and for $|X_1| \to \infty$ there is a cancellation so that the rhs of (\ref{half+-prime}) tends to the product of $-2^{(1/8)-1} /y_0^{(1/8)+1}$ and $6 ({\rm sign}X_1)/X_1^4$. Near $X_1 =0$, where the operator positions coincide, the behavior of the two contributions is quite different: While the two-point function is non-analytic as required by the OPE, the product of one-point functions is analytic and not related to the OPE. 

The behavior of $\langle \sigma \epsilon \rangle_{+-}^{(\rm cum)}$ given in Eqs. (\ref{half+-})-(\ref{half+-onepoint}) is entirely different from its counterpart for a uniform + boundary given in (\ref{+epssigprime}). On the other hand, for $x_1 \to -\infty$ with $x_2 -x_1 = y_0 X_{21}$ fixed, $\langle \sigma (x_1, y_0) \, \epsilon (x_2, y_0) \rangle_{+-}^{(\rm cum)}$ must approach the form of (\ref{+epssigprime}). This happens in an interesting way as shown \red{in FIG 1.}

The expression of the response function for {\it arbitrary} positions of $\sigma$ and $\epsilon$  follows from Eq. (4.3) in Ref. \cite{BX} and can be written as \cite{complex}
\begin{eqnarray} \label{se+-cum}
\langle \sigma (x_1 , y_1) \epsilon (x_2 , y_2) \rangle_{+-}^{\rm (cum)} = - {2^{(1/8)-1} \over y_1^{1/8} \, y_2} \times {1 \over \, |z_1| \, |z_2|^2} && \Biggl[{A+B+C \over |z_1 - z_2| \, |z_1 - \bar{z}_2|} + \nonumber \\
&& - \bigl(|z_2|^2 - z_2^2 - \bar{z}_2^2 \bigr) \, {z_1 +\bar{z}_1 \over 2} \Biggr]
\end{eqnarray}
where
\begin{eqnarray} \label{seA}
A&=&(|z_1|^2 +|z_2|^2) \, \bigl(|z_2|^2 -z_2^2 -\bar{z}_2^2 \bigr) {z_1 +\bar{z}_1 \over 2} \nonumber \\
B&=& {1 \over 2} \bigl( z_1 - \bar{z}_1 \bigr)^2 \, |z_2|^2 \, {z_2 +\bar{z}_2 \over 2} \nonumber \\
C&=& |z_1|^2 \, \bigl( z_2^3 + \bar{z}_2^3 \bigr) \, .
\end{eqnarray}
and the second term in the square bracket in (\ref{se+-cum}) arises from subtracting the product of one-point functions. Along the line $y = y_0$ this yields the expression
\begin{eqnarray} \label{se+-cumy0}
&&R_{+-} (X_1,X_2) \equiv \bigl( y_0^{(1/8)+1} \big/ 2^{(1/8)-1} \bigr) \, \langle \sigma (x_1 , y_0) \epsilon (x_2 , y_0) \rangle_{+-}^{\rm (cum)}=   \\
&& ={1 \over \, \sqrt{X_1^2 +1} \, (X_2^2 +1)} \; \Biggl\{{ X_1 (X_2^2 - 3) \bigl[X_1^2 + X_2^2 -2 X_1 X_2 + 2  \bigr] + 8 X_2 \over |X_1 - X_2| \, \sqrt{(X_1 - X_2)^2 +4}} - \nonumber \\
&& \qquad \qquad \qquad \qquad \qquad \qquad - X_1 \bigl(X_2^2 - 3 \bigr) \Biggr\} \, , \qquad X_1 \equiv x_1 /y_0 \, , \; X_2 \equiv x_2 /y_0 \,   \nonumber
\end{eqnarray}
on which FIG. 1 is based. The rhs of (\ref{se+-cumy0}) obeys the antisymmetry (\ref{+-symmetry}) and in the limit $X_1 \to - \infty$ with $X_2 - X_1$ fixed it reduces to the expression (\ref{+epssigprime}) for a uniform + boundary.

\section{Boundary condition $-+-$} \label{hp-+-}

Here we consider the two point function $\langle \sigma \epsilon \rangle_{-+-}$ and its cumulant in the upper half $h=g+ij$ plane with boundary conditions $-$ for $g< -1$, + for $-1<g<1$, and  $-$ for $g>1$. This function is symmetric about the imaginary axis $g=0$, i.e.,
\begin{eqnarray} \label{symm}
\langle \sigma (g_1, j_1) \, \epsilon (g_2, j_2) \rangle_{+-}^{(\rm cum)} =  \langle \sigma (-g_1, j_1) \, \epsilon (-g_2, j_2) \rangle_{+-}^{(\rm cum)}
\end{eqnarray}
which should be compared with the antisymmetry (\ref{+-symmetry}) for the $+-$ boundary.

The present two point function $\langle \sigma \epsilon \rangle_{-+-}$ in the $h$ plane is related to the two point function $\langle \sigma \epsilon \rangle_{+-}$ in the $z$ plane contained in Eqs. (\ref{se+-cum}), (\ref{seA})  by means of the Möbius transformation
\begin{eqnarray} \label{Möb}
z(h)= {h-1 \over h+1} \, , \quad {dz \over dh} = {2 \over (h+1)^2}.
\end{eqnarray}
The result is
\begin{eqnarray} \label{se-+-}
\langle \sigma (g_1 , j_1) \epsilon (g_2 , j_2) \rangle_{-+-} = - {1 \over 2 j_2} \, \Bigl( {2 \over j_1} \Bigr)^{1/8}  \times {a + b + c \over 4 |h_1 - h_2| \, |h_1 - \bar{h}_2|} 
\end{eqnarray}
where
\begin{eqnarray} \label{sealpha}
a&=&{|h_1|^2 - 1 \over |h_1^2 - 1|} \, \bigl[ |h_1 -1|^2 \, |h_2 +1|^2 + (h_1 \leftrightarrow h_2) \,  \bigr] \, \bigl( 1-2 \cos (2 \Phi_2) \bigr) \nonumber \\
b&=& {2 (h_1 - \bar{h}_1)^2  \, (|h_2|^2 - 1) \over |h_1^2 - 1|} \nonumber \\
c&=& |h_1^2 - 1| \, |h_2^2 - 1| 2 \cos (3 \Phi_2) \, . 
\end{eqnarray}
Here
\begin{eqnarray} \label{sePhi}
\Phi_2 \equiv {\rm arg}(h_2 -1) - {\rm arg}(h_2 +1) \, ,
\end{eqnarray}
and $\cos (2 \Phi_2)$ and $\cos (3 \Phi_2)$ arise from the rewriting 
\begin{eqnarray} \label{sePhiprime}
(h_2 -1)^2 \, (\bar{h}_2 +1)^2 + {\rm cc} =  |h_2^2 - 1|^2 \, 2 \cos (2 \Phi_2) \nonumber \\ 
{(h_2 -1)^2 \, (\bar{h}_2 +1)^2 \over (h_2 +1) \, (\bar{h}_2 -1)} + {\rm cc}=  |h_2^2 - 1| \, 2 \cos (3 \Phi_2)  
\end{eqnarray}
of terms that appear in the course of the transformation.

The expression for the cumulant $\langle \sigma \epsilon \rangle_{-+-}^{(\rm cum)} \equiv \langle \sigma \epsilon \rangle_{-+-} - \langle \sigma \rangle_{-+-} \times \langle \epsilon \rangle_{-+-}$ follows from Eqs. (\ref{se-+-})-(\ref{sePhiprime}) and the forms
\begin{eqnarray} \label{1sig-+-}
\langle \sigma (g, j) \rangle_{-+-} = - \Bigl( {2 \over j} \Bigr)^{1/8} \times {C} \, , \quad \langle \epsilon (g, j) \rangle_{-+-} = - {1 \over 2j} \, (4 {C}^2 - 3) \, , \quad {C} \equiv  {|h|^2 -1 \over |h^2 -1|}
\end{eqnarray}
of the one-point functions or profiles of $\sigma$ and $\epsilon$ in the upper $-+-$ half plane. They follow from their counterparts in the $+-$ plane given in Eq. (4.1) in Ref. \cite{BX}, see Eqs. (2.14) and (2.23) in Ref. \cite{E23}. The zero line with vanishing $\langle \sigma \rangle_{-+-}$ is the upper half unit circle which is the preimage of the imaginary axis in the $+-$ plane. In particular, the point $h=i$ on the zero line is the preimage of $z=i$. 

On mirror imaging both points about the imaginary axis in the $z$ plane each of the terms $(A,B,C)$ in (\ref{seA}) is antisymmetric while each of the terms $(a,b,c)$ in(\ref{sealpha}) is symmetric in the $h$ plane. Together with the symmetry of the corresponding denominators and the profiles in Eq. (\ref{1sig-+-}) this leads to the antisymmetry (\ref{+-symmetry}) and symmetry (\ref{symm})  of the cumulants in the $z$ and $h$ plane, respectively. We note that $(A,B,C)=(a,b,c)\times |h_1 -1||h_2 -1|^2 |h_1+1|^{-3} |h_2 +1|^{-4}$ on using the relation (\ref{Möb}).

Consider now the cases in which one of the two operators $\sigma$ and $\epsilon$ is located on a point of the zero line. Here we choose the point $h=i$.

\paragraph{$h_1 =i$}: Here Eqs. (\ref{se-+-})-(\ref{sePhiprime}) imply
\begin{eqnarray} \label{h1equi}
\langle \sigma (g_1 =0, j_1 =1) \, \epsilon (g_2 , j_2) \rangle_{-+-} = 2^{(1/8)+3} \, j_2 \, {|h_2|^2-1 \over |h_2^2 +1| \, |h_2^2 -1|^2 } \, 
\end{eqnarray}
and we note the corresponding form
\begin{eqnarray} \label{h1iT2}
\langle \sigma (g_1 =0 , j_1 =1) \, T(h) \rangle_{-+-} = 2^{1/8} \, {2 \over h^4 -1}
\end{eqnarray}
of $\langle \sigma \, T(h) \rangle_{-+-}$ that follows from subjecting (\ref{Tsigma}) with $y_0 =1$ to the transformation (\ref{Möb}). The near-boundary behavior $j_2 \to 0$ of (\ref{h1equi}) is given by $2^{(1/8)+3} \, j_2 /(g_2^4 -1 )$ which equals the product of $4 j_2$ and (\ref{h1iT2}) for $h=g_2$, as predicted by the BOE \cite{BOE}. We also note the expansion of (\ref{h1equi}) for small $|h_2 -h_1| \equiv |h_{21}| =  \sqrt{g_2^2 + (j_2 -1)^2} $,
\begin{eqnarray} \label{h1equiprime}
\langle \sigma (g_1 =0 , j_1 =1) \, \epsilon (g_2 , j_2) \rangle_{-+-} \to 	2^{(1/8)+1} {1 \over |h_{21} | } \times \nonumber \\
\times \Bigl\{ (j_2 -1) + \Bigl[ {1 \over 2}  g_2^2  - (j_2 -1)^2 \Bigr] - {7 \over 8} \, g_2^2 \, (j_2 -1)  + O \bigl( |h_{21} |^4 \bigr) \Bigr\} \, .
\end{eqnarray}
The first, second, and third term in the curly bracket of (\ref{h1equiprime}) is consistent with the  OPE-expressions given in (\ref{next}), (\ref{derivsig-+-}), and (\ref{OPE3-+-}), respectively. For $g_2 =0$ Eq. (\ref{h1equi}) takes the simple form
\begin{eqnarray} \label{h1equiprimeprime}
\langle \sigma (g_1 =0, j_1 =1) \, \epsilon (g_2 =0, j_2) \rangle_{-+-} = 2^{(1/8)+3} \, {j_2 \, {\rm sign }(j_2 -1) \over (j_2^2 +1)^2} 
\end{eqnarray}
displaying the remarkable behavior
\begin{eqnarray} \label{g1equg2}
\langle \sigma (g_1 =0, j_1 =1) \, \epsilon (g_2 =0, j_2) \rangle_{-+-} \to  2^{(1/8)+1} \, \bigl[-4 j_2 \, , \, {\rm sign}(j_2-1) -|j_2 -1| \, , \, 4j_2^{-3} \bigr]
\end{eqnarray}
for $\bigl[j_2 \to 0 \, , \, |j_2 -1| \ll 1 \, , \, j_2 \gg 1 \bigr]$ with a discontinuity at $j_2 =1$ and a prominent minimum at $j_2= 0.58$ where (\ref{h1equiprimeprime}) takes the value $-2^{(1/8)+1} \times 1.3$. Unlike Eq. (\ref{half+-}) where the antisymmetry (\ref{+-symmetry}) allows only contributions odd in $X_2$, in (\ref{h1equiprimeprime}) there is no symmetry in $j$-direction about the zero $j_1 =1$ of $\langle \sigma \rangle_{-+-}$ and both odd and even powers in $j_2 -1$ are present in the expansion which follows from (\ref{h1equiprime}) or (\ref{si-+-})-(\ref{OPE3-+-}). In the last paragraph of section \ref{uniform} and in Eq. (\ref{uni}) ff. we encountered this same phenomenon along the vertical $y$-axis of the upper half plane with a uniform + boundary.  

\paragraph{$h_2 =i$:}Here Eqs. (\ref{se-+-})-(\ref{sePhiprime}) imply
\begin{eqnarray} \label{h2equi}
\langle \sigma (g_1 , j_1) \, \epsilon (g_2 =0 , j_2 =1) \rangle_{-+-} = -{3 \over 2} \, \Bigl( {2 \over j_1} \Bigr)^{1/8} \, {|h_1|^4 -1 \over |h_1^4 -1|}  \, .
\end{eqnarray}
Eqs. (\ref{h1equi}) and (\ref{h2equi}) are invariant against $h_2 \to - \bar{h}_2$ and $h_1 \to - \bar{h}_1$, respectively, reflecting the mirror symmetry (\ref{symm}). From Eq. (\ref{h2equi}) follows the short distance behavior. 
\begin{eqnarray} \label{h2equishort}
&&\langle \sigma (g_1 =0 , j_1) \, \epsilon (g_2 =0 , j_2 =1) \rangle_{-+-} \equiv -{3 \over 4} \, 2^{(1/8)+1} \, j_1^{-1/8} \, {\rm sign}(j_1 -1) \, \to  \nonumber \\ 
&&\qquad \qquad  \to -{3 \over 4} \, 2^{(1/8)+1} \, \bigl[  {\rm sign}(j_1 -1) - (1/8) \, |j_1 -1| +O\bigl( (j_1 -1)^2 \bigr) \bigr]
\end{eqnarray}
which should be compared with the short distance behavior in Eq. (\ref{g1equg2}) and which is consistent with the OPE-expression in (\ref{h2OPEshort}). Subtracting the expanded product of one-point functions given in (\ref{1sig-+-}) then yields 
\begin{eqnarray} \label{h2equishortcum}
&&\langle \sigma (g_1 =0 , j_1) \, \epsilon (g_2 =0 , j_2 =1) \rangle_{-+-}^{(\rm cum)}  \to \nonumber \\
&& \qquad \to -{3 \over 4} \, 2^{(1/8)+1} \, \bigl[  {\rm sign}(j_1 -1) - (1/8) \, |j_1 -1| -(j_1 -1) +O\bigl( (j_1 -1)^2 \bigr) \bigr]
\end{eqnarray}
for the cumulant.

\subsection{Behavior of $\langle \sigma \, \epsilon \rangle_{-+-}^{\rm cum}$ for arbitrary points on the symmetry axis} \label{g1 =g2 =0}

For both $\sigma$ and $\epsilon$ positioned at arbitrary points on the imaginary axis of the $h$ plane, i.e., for $g_1 =g_2 =0$, Eqs. (\ref{se-+-}) ff. yield for the cumulant
\begin{eqnarray} \label{12imag}
\qquad \langle \sigma (g_1 =0 , j_1) \, \epsilon (g_2 =0 , j_2) \rangle_{-+-}^{(\rm cum)} &=& -{1 \over 2j_2} \Bigl( {2 \over j_1} \Bigr)^{1/8} \, {\cal A}(j_1^2 \, , \, j_2^2) \, , \, \\
{\cal A}(p,q) \, \equiv \, {2 \over |p-q| (p+1)(q+1)^2} &\times& \Bigl[ q(1+15p-14q-2pq+q^2-pq^2)\, , \, \nonumber \\
&& \quad p(1-p+2q+14pq-15q^2-pq^2) \Bigr] \nonumber
\end{eqnarray}
for $\bigl[ q<p \, , \, p<q \bigr]$. The antisymmetry about the imaginary axis in the $z$ plane addressed in Sec. \ref{hp+-} translates along the present imaginary axis in the $h$ plane to an inversion antisymmetry of ${\cal A}$ about $h=i$ so that
\begin{eqnarray} \label{inversion}
{\cal A}(p^{-1} \, , \, q^{-1}) \, = \, - {\cal A}(p \, , \, q) \, .
\end{eqnarray}

The special cases
\begin{eqnarray} \label{jumpsA} 
{\cal A}(1,q) = -{16 q \over (q+1)^2}\times {\rm sign} (q-1) \, , \quad {\cal A}(p,1) = 3 \, \Bigl[{\rm sign} (p-1) - {p-1 \over p+1} \Bigr] \, 
\end{eqnarray}
serve to rederive the cumulant-expansions (\ref{g1equg2}), (\ref{h2equishortcum}) for $(j_1=1, \, j_2 \to 1) \, , \, (j_2 =1, \, j_1 \to 1)$. The universal ratio $-4/3$ mentioned in the Introduction after Eq. (\ref{x2zero}) immediately follows from the leading terms for $q \to 1$, $p \to 1$ in (\ref{jumpsA}).

For later use we note the limiting behaviors
\begin{eqnarray} \label{limiteps}
\langle \sigma (g_1 =0 , j_1) \, \epsilon (g_2 =0 , j_2) \rangle_{-+-}^{(\rm cum)} \to 	\Bigl( {2 \over j_1} \Bigr)^{1/8} \, {1 \over j_1^2 +1} \Bigl\{- {j_2 \, (15 j_1^2 +1) \over j_1^2} \, , \, {j_1^2 \, (j_1^2 +15) \over j_2^3}  \Bigr\}
\end{eqnarray}
for $\{ j_2 \to 0 \, , \, j_2 \to \infty \}$ and
\begin{eqnarray} \label{limitsig}
\langle \sigma (g_1 =0 , j_1) \, \epsilon (g_2 =0 , j_2) \rangle_{-+-}^{(\rm cum)} &\to& 	\Bigl( {2 \over j_1} \Bigr)^{1/8} \, {1 \over (j_2^2 +1)^2} \times \nonumber \\
&&\times\Bigl\{ {j_1^2 \, (15 j_2^4 -2 j_2^2 - 1) \over j_2^3} \, , \, {j_2 \, (j_2^4 +2 j_2^2 - 15) \over j_1^2}  \Bigr\}
\end{eqnarray}
for $\{ j_1 \to 0 \, , \, j_1 \to \infty \}$.

Here it is useful to make contact with the cumulants containing the stress tensor discussed in Appendix \ref{stresscum}. For $j_1$ arbitrary fixed and $j_2$ tending either to $0$ or to $\infty$ one finds 
\begin{eqnarray} \label{limits}
\langle \sigma (g_1 =0 , j_1) \, \epsilon (g_2 =0 , j_2) \rangle_{-+-}^{(\rm cum)} \, \big/ \, \langle \sigma (g_1 =0 , j_1) \, T(h=ij_2) \rangle_{-+-}^{(\rm cum)} \, \to \, 4j_2 \,   
\end{eqnarray}
while for $j_2$ arbitrary fixed and $j_1$ tending either to $0$ or to $\infty$
\begin{eqnarray} \label{limitsprime}
\langle \sigma (g_1 =0 , j_1) \, \epsilon (g_2 =0 , j_2) \rangle_{-+-}^{(\rm cum)} \, \big/ \, \langle T(h=ij_1) \, \epsilon (g_2 =0 , j_2) \, \rangle_{-+-}^{(\rm cum)} \, \to \nonumber \\
\to \, \{ - \, , +  \} \, \Bigl( {2 \over j_1} \Bigr)^{1/8} \, j_1^2 \, .  
\end{eqnarray}
This follows by comparing Eqs. (\ref{limiteps}) with the corresponding limiting behaviors of $\langle \sigma \, T \rangle_{-+-}^{(\rm cum)}$ given in Eqs. (\ref{sT}) and, likewise, (\ref{limitsprime}) with (\ref{Te}). For $j_2 \to 0$ and $j_1 \to 0$ where $\epsilon$ and $\sigma$, respectively, approach the boundary the relations confirm the BOE \cite{BOE}. 

In our discussion of a triangle in Sec. \ref{equitri} we shall use the properties (\ref{limits}) and (\ref{limitsprime}) for $j_2 \to \infty$ and $j_1 \to \infty$ to show that on approaching a corner of the triangle the behavior is consistent with the ``corner operator expansion'' COE \cite{COE}.


\section{Equilateral triangle with $-+-$ boundary condition} \label{equitri}

Here we consider the critical behavior inside an equilateral triangle in the $z=x+iy$ plane with side length ${\cal W}$, with corners at 
\begin{eqnarray} \label{trihalfprime} 
z = z_{\rm A} = - {\cal W}/2, \, z_{\rm B} = {\cal W} /2, \, z_{\rm C} = i (\sqrt{3} /2) {\cal W} =i y_{\rm C} \, ,
\end{eqnarray}
and with boundary conditions + along the horizontal AB side and $-$ along the CA and CB sides. Correlation functions can be related to those in the upper half $h=g+ij$ plane described in section \ref{hp-+-} by means of the conformal transformation $h(Z)$ in Appendix \ref{trihalf}. Here one uses the dimensionless variable $Z=X+iY$ defined in Eq. (\ref{trhalf}) which measures the position $z$ conveniently in terms of the side length ${\cal W}$ or the height $y_{\rm C} = (\sqrt{3} /2) {\cal W}$ of the triangle. The relation for our $\sigma \times \epsilon$ cumulant then reads 
\begin{eqnarray} \label{sigepstrafo} 
&& \qquad \langle \sigma (x_1, y_1) \, \epsilon (x_2,y_2) \rangle_{\rm triangle}^{(\rm cum)} = \Bigl( { Y_{\rm C} \over y_{\rm C}} \Bigr)^{(1/8)+1} \times  \nonumber \\
&&\times  |S(Z_1)|^{1/8} \, |S(Z_2)| \, \langle \sigma (g_1 , j_1) \epsilon (g_2 , j_2) \rangle_{-+-}^{\rm (cum)} \, , \quad S(Z) \equiv {dh \over dZ} \, .
\end{eqnarray}
Here $Y_{\rm C} = 3.196284004$ is the $Y$ argument of the upper corner, see Eq. (\ref{twomid}).

It is instructive to consider the behavior along the vertical midline $Z=iY, \, 0<Y< Y_C$ of the triangle which is mapped to the midline $h=ij,\, 0<j< \infty$ of the half plane system in section \ref{hp-+-}. The explicit expressions of $S(iY)=|S(iY)|$ and $j(Y)$ are given in Eqs. (\ref{rescmid'}) and (\ref{trafomid}), respectively. We note the form
\begin{eqnarray} \label{sigtri} 
\langle \sigma (x=0, \, y) \rangle_{\rm triangle} \, = \, \Bigl( {Y_{\rm C} \over y_{\rm C}} \Bigr)^{1/8} \times S(iY)^{1/8} \times \Bigl( {2 \over j} \Bigr)^{1/8} \times {1-j^2 \over 1+j^2} 
\end{eqnarray}
of the one-point function $\langle \sigma \rangle_{\rm triangle}$ along the midline which follows from Eq. (\ref{1sig-+-}). It vanishes for $Y=Y_0 \equiv 0.74326$ since this corresponds to $j=1$, see Eq. (\ref{twomid}), and it is positive and negative for $Y<Y_0$ and $Y>Y_0$, respectively. 

First we discuss the behavior of the response function (\ref{sigepstrafo}) when $\epsilon$ or $\sigma$ approach the base line or the upper corner of the triangle. Boundary operator expansions serve for a deeper understanding what happens when a bulk operator approaches a flat boundary or corner \cite{Diehl,CardBoundaryCritPhen,BE21,E23}. Like in Sec. \ref{hp-+-} and Appendix \ref{stresscum} we relate the average (\ref{sigepstrafo}) of $\sigma \times \epsilon$ along the midline to the corresponding averages of $\sigma \times T$ and $T \times \epsilon$. These read
\begin{eqnarray} \label{sigTtrafo} 
&&\langle \sigma (x_1 =0, y_1) \, T(iy_2) \rangle_{\rm triangle}^{(\rm cum)} = \Bigl( {Y_{\rm C} \over y_{\rm C} } \Bigr)^{(1/8)+2} \times  \nonumber \\
&& \qquad \times  S(i Y_1)^{1/8} \, S^2 (i Y_2) \, \langle \sigma (g_1 =0 , j_1)  T(i j_2) \rangle_{-+-}^{\rm (cum)} \, , 
\end{eqnarray}
and
\begin{eqnarray} \label{Tepstrafo} 
&&\langle  T(iy_1)  \epsilon (x_2 =0, y_2) \,\rangle_{\rm triangle}^{(\rm cum)} = \Bigl( {Y_{\rm C} \over y_{\rm C} } \Bigr)^{2+1} \times  \nonumber \\
&& \qquad \times  S(i Y_1)^2 \, S(i Y_2) \, \langle T(i j_1) \epsilon (g_2 =0 , j_2)   \rangle_{-+-}^{\rm (cum)} \, , 
\end{eqnarray}
respectively, and we find the ratios
\begin{eqnarray} \label{nearcorner} 
{\langle \sigma (x_1 =0, y_1) \, \epsilon(x_2 =0, y_2) \rangle_{\rm triangle}^{(\rm cum)} \over \langle \sigma (x_1 =0, y_1) \, T(iy_2) \rangle_{\rm triangle}^{(\rm cum)}} \to {y_{\rm C} \over Y_{\rm C} } \, {4 j (Y_2) \over R(i Y_2)} \, \to \, 4 \bigl\{y_2 \, , \, (y_{\rm C} -y_2)/3 \bigr\} \, .
\end{eqnarray}
and 
\begin{eqnarray} \label{nearcornerprime} 
{\langle \sigma (x_1 =0, y_1) \, \epsilon(x_2 =0, y_2) \rangle_{\rm triangle}^{(\rm cum)} \over \langle T(iy_1) \, \epsilon (x_2 =0, y_2) \rangle_{\rm triangle}^{(\rm cum)}} \, &\to& \{- \, , \, + \} \, 2^{1/8} %
\, \Bigl( {y_{\rm C} \over Y_{\rm C}} \, {j (Y_1) \over R(i Y_1)} \Bigr)^{2-(1/8)} \, \to \nonumber \\
&\to& \, 2^{1/8} \, \bigl\{- y_1^{2-(1/8)} \, , \, [ (y_{\rm C} -y_1)/3 ]^{2-(1/8)} \bigr\} \, 
\end{eqnarray}
in the limits $\{y_2 \to 0 \, , \, y_2 \to y_{\rm C} \}$ and $\{y_1 \to 0 \, , \, y_1 \to y_{\rm C} \}$, respectively. To derive (\ref{nearcorner}) and (\ref{nearcornerprime}) we have used Eqs. (\ref{limits}) and (\ref{limitsprime}) in the first steps  and Eqs. (\ref{jandRnearbase}), (\ref{jandRnearC}) in the second steps. That the ratios in (\ref{nearcorner}) and (\ref{nearcornerprime}) are {\it independent} of $y_1$ and $y_2$, respectively, and have the simple forms given on their right hand sides are important consequences of the BOE explained in \cite{BOE} and of the COE \cite{COE} on using relations given in Eqs. (4.5) in Ref. \cite{E23}.  

Now we discuss the above limiting behaviors for the response function in the numerators of (\ref{nearcorner}) and (\ref{nearcornerprime})  which do depend on $y_1$ and $y_2$, respectively. Using the short notation
\begin{eqnarray} \label{seshort} 
\langle \sigma (x_1 =0, y_1) \, \epsilon(x_2 =0, y_2) \rangle_{\rm triangle}^{(\rm cum)} \equiv {\cal R} (y_1 \, , \, y_2)
\end{eqnarray}
for the response function they read
\begin{eqnarray} \label{y2to0} 
{\cal R} (y_1 \, , \, y_2) &\to& \Bigl( { Y_{\rm C} \over y_{\rm C}} \Bigr)^{(1/8)+1} \, 3^{3/2} \times (Y_2 /4) \times {\cal B}_2 (Y_1) \, , \nonumber \\
{\cal B}_2 (Y_1)&=& - \bigl[3^{3/4} (1+j_1^2)^{2/3} /j_1 \bigr]^{1/8} \, {15j_1^2 +1 \over j_1^2 (j_1^2 +1)}
\end{eqnarray}
and
\begin{eqnarray} \label{y2toyC} 
{\cal R} (y_1 \, , \, y_2) &\to& \Bigl( { Y_{\rm C} \over y_{\rm C}} \Bigr)^{(1/8)+1} \, {1 \over 2} 3^{-1/2} \times (\delta_2/2)^5 \times {\cal C}_2 (Y_1) \, , \nonumber \\ 
{\cal C_2} (Y_1)&=&\bigl[3^{3/4} (1+j_1^2)^{2/3} /j_1 \bigr]^{1/8} \, {j_1^2 (j_1^2 +15) \over j_1^2 +1}
\end{eqnarray}
for $y_2 \to 0$ and $y_2 \to y_{\rm C}$, respectively. For $y_1 \to 0$ and $y_1 \to y_{\rm C}$ the result is 
\begin{eqnarray} \label{y1to0} 
{\cal R} (y_1 \, , \, y_2) &\to& \Bigl( { Y_{\rm C} \over y_{\rm C}} \Bigr)^{(1/8)+1} \, 3^{3/2} \times (Y_1 /2)^{2-(1/8)} \times {\cal B}_1 (Y_2) \, , \nonumber \\
{\cal B}_1 (Y_2)&=& {1 \over 2} 3^{3/4} (1+j_2^2)^{2/3} \, {15j_2^4 - 2 j_2^2 -1 \over j_2^3 \, (j_2^2 +1)^2}
\end{eqnarray}
and 
\begin{eqnarray} \label{y1toyC} 
{\cal R} (y_1 \, , \, y_2) &\to& \Bigl( { Y_{\rm C} \over y_{\rm C}} \Bigr)^{(1/8)+1} \, 3^{-11/8} (\delta_1 /2)^{6-(1/8)} \times {\cal C}_1 (Y_2) \, , \nonumber \\ 
{\cal C_1} (Y_2)&=& {1 \over 2} \, 3^{3/4} (1+j_2^2)^{2/3} \, {j_2 (j_2^4 +2 j_2^2 -15) \over (j_2^2 +1)^2} \, ,
\end{eqnarray}
respectively. Here $j_1$ and $j_2$ means $j (Y_1)$ and $j (Y_2)$, as given in Eq. (\ref{trafomid}), and $\delta_1 \equiv Y_C -Y_1$, $\delta_2 \equiv Y_C -Y_2$. 

We note the interesting special cases in which one of the two operators $\sigma$ or $\epsilon$ is close to the base line while the other one is close to the corner. The corresponding results are  
\begin{eqnarray} \label{20,1C} 
{\cal R} (y_1 \, , \, y_2) &\to& - 5 \Bigl( {Y_{\rm C} \over y_{\rm C}} \Bigr)^8 \, y_2 \, (y_{\rm C} -y_1)^{6-(1/8)} \, 2^{-8+(1/8)} \, 3^{1+(1/8)} \, , \nonumber \\
&& \quad y_2 \to 0 \, , \, y_1 \to y_{\rm C}
\end{eqnarray}
and
\begin{eqnarray} \label{10,2C} 
{\cal R} (y_1 \, , \, y_2) &\to& 5 \Bigl( {Y_{\rm C} \over y_{\rm C}} \Bigr)^8 \, y_1^{2-(1/8)} \, (y_{\rm C} -y_2)^5 \, 2^{-8+(1/8)} \, 3^2 \, , \nonumber \\
&& \quad y_1 \to 0 \, , \, y_2 \to y_{\rm C} \, .
\end{eqnarray}
The first one follows from either putting $Y_1$ close to $Y_{\rm C}$ in ${\cal B}_2$ in Eq. (\ref{y2to0}) or from  putting $Y_2 \to 0$ in ${\cal C}_1$ in Eq. (\ref{y1toyC}) and the second one  from either putting $Y_1 \to 0$ in ${\cal C}_2$ in Eq. (\ref{y2toyC}) or from  putting $Y_2$ close to $Y_{\rm C}$ in ${\cal B}_1$ in Eq. (\ref{y1to0}).

For completeness we mention the special cases of (\ref{y2to0})-(\ref{y1toyC}) in which $y_1$ and $y_2$ are both near the base line or both near the corner: Putting $y_1 \ll y_{\rm C}$ in ${\cal B}_2$ and $y_{\rm C} -y_1 \ll y_{\rm C}$ in ${\cal C}_2$ yields
\begin{eqnarray} \label{purehalfplane} 
{\cal R} (y_1 \, , \, y_2) \to - \Bigl( {2 \over y_1} \Bigr)^{1/8} \, {y_2 \over y_1^2} \, , \quad y_2 \ll y_1 \ll y_{\rm C}
\end{eqnarray}
and
\begin{eqnarray} \label{purewedgeprime} 
{\cal R} (y_1 \, , \, y_2) &\to& 3^{(1/8)+1} \, \Bigl( {2 \over y_{\rm C} - y_1} \Bigr)^{1/8} \, {(y_{\rm C} -y_2)^5 \over (y_{\rm C} - y_1)^6} \, , \nonumber \\
&&y_{\rm C} -y_2 \ll y_{\rm C} -y_1 \ll y_{\rm C} \, , 
\end{eqnarray}
respectively. As expected Eqs. (\ref{purehalfplane}) and (\ref{purewedgeprime}) reproduce the simple results when the triangle degenerates to the infinitely extended base line and to the infinitely extended wedge, respectively, the point 2 being much closer to the basis and to the corner of the wedge, respectively, than point 1. These results follow from Eq. (\ref{+epssigprimeprime}) and its wedge transform. Likewise, for the cases in which point 1 is much closer than point 2 to the basis and the corner one invokes Eqs. (\ref{y1to0}) and (\ref{y1toyC}) with ${\cal B}_1 (Y_2 \to 0)$ and ${\cal C}_1 (Y_2 \to Y_{\rm C})$, respectively, which yields the expected results 
\begin{eqnarray} \label{purehalfplaneprime} 
{\cal R} (y_1 \, , \, y_2) \to - \Bigl( {2 \over y_1} \Bigr)^{1/8} \, {y_1^2 \over y_2^3} \, , \quad y_1 \ll y_2 \ll y_{\rm C}
\end{eqnarray}
and
\begin{eqnarray} \label{purewedge} 
{\cal R} (y_1 \, , \, y_2) &\to& 3^{(1/8)+1} \, \Bigl( {2 \over y_{\rm C} - y_1} \Bigr)^{1/8} \, {(y_{\rm C} -y_1)^6 \over (y_{\rm C} - y_2)^7} \, , \nonumber \\
&&y_{\rm C} -y_1 \ll y_{\rm C} -y_2 \ll y_{\rm C} \, . 
\end{eqnarray}

In Eqs. (\ref{shorttri1at0}) and (\ref{shorttri2at0}) below we discuss the short distance behavior of ${\cal R}$ for  the cases in which $\sigma$ or $\epsilon$ are located at the point $y = y_0 = 0.2325 \times y_{\rm C}$ of the midline where $\langle \sigma \rangle_{\rm triangle}$ vanishes \cite{center}, see Eq. (\ref{twomid}) ff. This point belongs to the zero line of our triangle that starts and ends at the corners $z_{\rm A}$ and $z_{\rm B}$ and crosses the midline at $z=i y_0$. 

Finally we consider the behavior of our response function on the vertical midline of the triangle, $\langle \sigma (x_1 =0, y_1) \, \epsilon(x_2 =0, y_2) \rangle_{\rm triangle}^{(\rm cum)} \equiv  {\cal R} (y_1 \, , \, y_2)$,  for {\it arbitrary} $y_2$ with $y_1$ fixed at several values and vice versa. This is shown in panels (a) and (b), respectively, of FIG. 2 in terms of the dimensionless response function $R_{\rm tri} (Y_1 \, , \, Y_2)$ which is defined by
\begin{eqnarray} \label{tildeR} 
{\cal R} (y_1 \, , \, y_2) = \Bigl( { Y_{\rm C} \over y_{\rm C}} \Bigr)^{(1/8)+1} \, R_{\rm tri} (Y_1 \, , \, Y_2) \, , \quad Y={Y_{\rm C} \over y_{\rm C}} \, y \, .
\end{eqnarray}
Since the midline of the triangle extends over the interval $0<y<y_{\rm C}$, the variables $Y_1$ and $Y_2$ extend along $0<Y<Y_{\rm C} \equiv 3.196$, see  Eq. (\ref{twomid}). The original order is positive and negative for $Y<Y_0$ and $Y>Y_0$, respectively, with $Y_0 \equiv 0.7432$, see Eq. (\ref{sigtri}). To calculate $R_{\rm tri}$ one uses (\ref{sigepstrafo}) that implies
\begin{eqnarray} \label{sigepstrafoprime} 
R_{\rm tri} (Y_1 \, , \, Y_2) = S(i Y_1)^{1/8} \, S(i Y_2) \, \langle \sigma (g_1 =0 , j_1) \, \epsilon (g_2 =0 , j_2) \rangle_{-+-}^{\rm (cum)} \big|_{j=j(Y)} \, 
\end{eqnarray}
where the expressions for the cumulant $\langle ... \rangle_{-+-}^{(\rm cum)}$, for $j(Y)$, and for $S(iY)$ are given in Eqs. (\ref{12imag}), (\ref{trafomid}), and (\ref{rescmid'}), respectively.

Eq. (\ref{sigepstrafoprime}) implies for $Y_1 = Y_0$ and for $Y_2 = Y_0$ the short distance behaviors up to linear order
\begin{eqnarray} \label{shorttri1at0} 
R_{\rm tri} (Y_1 =Y_0 \, , \, Y_2) \, \to \, \bigl( 2 S_0 \bigr)^{9/8} \times \Bigl[{\rm sign} (Y_2-Y_0) \, - \, {S_0 \over 3} \, |Y_2-Y_0| + O \big( (Y_2-Y_0)^2  \big) \Bigr]
\end{eqnarray}
and
\begin{eqnarray} \label{shorttri2at0} 
R_{\rm tri} (Y_1 \, , \, Y_2 =Y_0) &\to& - \bigl( 2 S_0 \bigr)^{9/8} \times {3 \over 4} \, \Bigl[{\rm sign} (Y_1-Y_0) \, - \nonumber \\
&& - {S_0 \over 3} \, \Bigl({|Y_1-Y_0| \over 8} + 3 \, (Y_1-Y_0) \Bigr) + O \bigl( (Y_1-Y_0)^2 \bigr)  \Bigr] \, .
\end{eqnarray}
Here the number $S_0$ is given by Eq. (\ref{closetoprime}) and we used the corresponding behaviors (\ref{g1equg2}) and (\ref{h2equishortcum}) in the $-+-$ half plane and the relations (\ref{closeto}). The universal ratio $-4/3$ between the upward and downward discontinuities is clearly visible. The tangents to the left and right of the discontinuities in panels (a) and (b) of FIG 2 are given by the linear order terms in Eqs. (\ref{shorttri1at0}) and (\ref{shorttri2at0}), respectively. In Sec. \ref{OPEtri} Eq. (\ref{shorttri1at0}) is confirmed by means of the operator-product expansion.

\section{${\cal W} \times {\cal W}$ square with vertical boundaries - and horizontal boundaries $+$} \label{-+-+square}

Consider a square with vertical boundaries - and horizontal boundaries $+$ with its center at the origin of the entire $z=x+iy$ plane, which is mirror-symmetric about the coordinate axes, implying for the two-point function $\langle \sigma  \, \epsilon  \rangle_{\rm SQ}$ for example that
\begin{eqnarray} \label{mirrormidline} 
\langle \sigma (x_1, y_1) \, \epsilon (x_2, y_2) \rangle_{\rm SQ}^{(\rm cum)} &=&  \langle \sigma (-x_1, y_1) \, \epsilon (-x_2, y_2) \rangle_{\rm SQ}^{(\rm cum)} \, ,
\nonumber \\
&=&  \langle \sigma (x_1, -y_1) \, \epsilon (x_2, -y_2) \rangle_{\rm SQ}^{(\rm cum)} \, .
\end{eqnarray}
Moreover, mirror-imaging about the diagonals interchanges boundary conditions $+ \leftrightarrow -$ which turns $\sigma \, , \, \epsilon$ into $- \sigma \, , \, \epsilon$. Thus 
\begin{eqnarray} \label{mirrordiagonal} 
\bigl[ \langle \sigma (x, y) \rangle_{\rm SQ} \bigr] \Big|_{x,y \to y,x} = - \langle \sigma (x, y) \rangle_{\rm SQ} \, , \quad \bigl[ \langle \epsilon (x, y) \rangle_{\rm SQ} \bigr] \Big|_{x,y \to y,x} =  \langle \epsilon (x, y) \rangle_{\rm SQ}
\nonumber \\
\big[ \langle \sigma (x_1, y_1) \, \epsilon (x_2, y_2) \rangle_{\rm SQ} \bigr] \Big|_{x_1,y_1;x_2,y_2 \to y_1,x_1;y_2,x_2}  = - \langle \sigma (x_1, y_1) \, \epsilon (x_2, y_2) \rangle_{\rm SQ}
\end{eqnarray}
and, in particular, $\langle \sigma \rangle_{\rm SQ}$ vanishes at the diagonals. The same symmetry relations apply to the two-point cumulants.

Putting either $\sigma$ or $\epsilon$ at the origin, where the two zero-line diagonals intersect,  and the other operator close to it, the leading behavior is given by
\begin{eqnarray} \label{originsigmaepsilon} 
\langle \sigma (x_1 =0, y_1 =0) \, \epsilon (x_2, y_2) \rangle_{\rm SQ} &\to&  {16 \over 3} \Lambda^{-(1/8)-2} \, 2^{1/4} \times {x_2^2 - y_2^2 \over \sqrt{x_2^2 + y_2^2}} \nonumber \\
\langle \sigma (x_1 , y_1) \, \epsilon (x_2 =0, y_2 =0) \rangle_{\rm SQ} &\to& - {5 \over 3} \Lambda^{-(1/8)-2} \, 2^{1/4} \times {x_1^2 -y_1^2 \over \sqrt{x_1^2 +y_1^2}}\,
\end{eqnarray}
which is different from the cases in Secs. \ref{hp+-}-\ref{equitri}. Along the midlines the non-analyticities are not discontinuities but rather the symmetric cusps
\begin{eqnarray} \label{originsigma} 
\langle \sigma (x_1 =0, y_1 =0) \, \epsilon (x_2, y_2 =0) \rangle_{\rm SQ} &\to&  {16 \over 3} \Lambda^{-(1/8)-2} \, 2^{1/4} \,\times x_2 \; {\rm sign}(x_2) \, , \nonumber \\
\langle \sigma (x_1 =0, y_1 =0) \, \epsilon (x_2 =0, y_2) \rangle_{\rm SQ} &\to& - {16 \over 3} \Lambda^{-(1/8)-2} \, 2^{1/4} \times y_2 \; {\rm sign}(y_2)
\end{eqnarray}
and
\begin{eqnarray} \label{originepsilon} 
\langle \sigma (x_1 , y_1 =0) \, \epsilon (x_2 =0, y_2 =0) \rangle_{\rm SQ} &\to& - {5 \over 3} \Lambda^{-(1/8)-2} \, 2^{1/4} \times x_1 \; {\rm sign}(x_1) \, , \nonumber \\
\langle \sigma (x_1 =0, y_1 ) \, \epsilon (x_2 =0, y_2 =0) \rangle_{\rm SQ} &\to&  {5 \over 3} \Lambda^{-(1/8)-2} \, 2^{1/4} \, y_1 \times {\rm sign}(y_1) \, .
\end{eqnarray}
Still qualitative features found in Secs. \ref{hp+-}-\ref{equitri} remain. E.g., Eq. (\ref{originepsilon}) tells us that the order is strengthend by increasing the disorder at the center of our square.

In the above equations $\Lambda^{-1} = \bigl({\bf K}(1/\sqrt{2} \bigr) \, / {\cal W})$ with ${\bf K}(1/\sqrt{2}) = 1.854$ the complete elliptic integral. These results apply when the nonvanishing coordinate is much smaller than ${\cal W}$. They are consistent with the above symmetry relations and, as discussed in Sec.\ref{OPEsq}, they follow via the ${\cal a}$, ${\cal b}$, and ${\cal c}$ terms in the OPE (\ref{ope3}) from the \red{form} 
\begin{eqnarray} \label{sigSQ} 
\langle \sigma (x_1 , y_1 ) \rangle_{\rm SQ} \to - \Lambda^{-(1/8)-2} \, 2^{1/4} \, 2(x_1^2 - y_1^2) \, = \, - \Lambda^{-(1/8)-2} \,  2^{1/4} \, (z_1^2 + \bar{z}_1^2) \, ,
\end{eqnarray}
of the one-point function near the center,  which is derived in Appendix \ref{SQU}. Eqs. (\ref{originsigmaepsilon})-(\ref{originepsilon}) apply not only to the two-point function but also to the cumulant $\langle \sigma  \, \epsilon \rangle_{\rm SQ}^{\rm (cum)}$ since their difference,
\begin{eqnarray} \label{prodsigepsSQ} 
\langle \sigma (x_1 , y_1) \rangle_{\rm SQ} \times \langle \epsilon (x_2 , y_2 ) \rangle_{\rm SQ} \to - {10 \over 3} \Lambda^{-(1/8)-3} \, 2^{1/4} \, (x_1^2 -y_1^2) \, ,
\end{eqnarray}
is smaller by a factor $x / \Lambda$ or $y / \Lambda$. Eq. (\ref{prodsigepsSQ}) arises from (\ref{sigSQ}) and the form $\langle \epsilon (x_2 , y_2 ) \rangle_{\rm SQ} \to (5/3) \, \Lambda^{-1}$ of the energy density profile near the center.


\section{SUMMARY AND CONCLUDING REMARKS} \label{summ}

We consider the two-dimensional critical Ising model with mixed boundary conditions and ask how local ordering imposed at point ${\bf r}_1$ affects the disorder at another point ${\bf r}_2$  and vice versa. The answer is contained in the universal cumulant response function $\langle \sigma ({\bf r}_1 )\epsilon ({\bf r}_2)\rangle^{(\rm cum)}$, where $\sigma$ and $\epsilon$ are the density operators of the order parameter and energy. Making use of the OPE for $\sigma\times\epsilon$ and exact results, we study the response function of systems in the upper half plane with (i) a uniform boundary $+$ of fixed up spins, (ii) a mixed boundary $+-$ of fixed up and down spins on the negative and positive boundary line, respectively, and (iii) a $-+-$ boundary consisting of a finite segment of up spins between two semi-infinite segments of down spins. We also consider two finite systems: (iv) an equilateral triangle with up spins on one edge, the horizontal base line of the triangle, and down spins on the other two edges, and  (v) a square with up spins on the horizontal edges and down spins on the vertical edges. The mixed boundaries in (ii)-(v) generate zero lines along which the order-parameter profile $\langle \sigma \rangle$ vanishes.

FIGs. 1 and 2 show the remarkable behavior of the response function, associated with zero lines, as ${\bf r}_1$ and ${\bf r}_2$ vary along a line parallel to the boundary of the half plane (ii) and along the vertical midline of the triangle (iv). 

The response function $\langle \sigma\epsilon\rangle^{(\rm cum)}$ for (ii) is known exactly \cite{BX}, and transforming it conformally leads to exact expressions for (iii) and (iv). Despite this, some of the interesting implications seem to have been overlooked.  We have shown the utility of the OPE in analyzing the behavior for small $|{\bf r}_{12}|$,  an approach which is not limited to the Ising model but applicable to a broader class of systems.
	
The OPE'S in Eqs.~(\ref{selfOPE}) and (\ref{next}) imply that in {\it leading} order, $\langle \sigma\epsilon\rangle^{(\rm cum)}$ at short distances $|{\bf r}_{12}|$ depends on local properties of the order-parameter density $\langle \sigma \rangle$. This is in contrast to the correlation functions $\langle \sigma\sigma\rangle^{(\rm cum)}$ and  $\langle\epsilon\epsilon\rangle^{(\rm cum)}$, where the dependence on the corresponding symmetry-allowed profiles $\langle \epsilon \rangle$ or $\langle T \rangle$ only appears in higher order \cite{spinreverse}. For nonvanishing  $\langle \sigma \rangle$, the leading singularity of $\langle \sigma\epsilon \rangle^{(\rm{cum})}$ has the form of a power law divergence with magnitude proportional to $\langle \sigma \rangle$ but with the opposite sign, see Eq. (\ref{next}). On placing one of the two operators on a zero line of  $\langle \sigma \rangle$ and crossing the line with the other operator, the leading short-distance singularity is milder, having the form of a discontinuity, a cusp, etc., depending on the lowest nonvanishing derivative of $\langle \sigma \rangle$. For systems (ii), (iii), and (iv), the singularity is a discontinuity, while at the center of the square (v), where the first derivatives vanish, it is a cusp.  The ratio of  the two discontinuities, when one of the two operators $\sigma$ or $\epsilon$ is placed on a point of the zero line while the other one crosses it, is a universal number. All this is a consequence of the operator-product expansion (\ref{next}) and its extension to higher order in Appendix A.
	
We also analyze the behavior of the response function as one of the two operators $\sigma$ or $\epsilon$ approaches a flat boundary or a corner, making use of the boundary- operator expansion or corner-operator expansion, respectively. As the operator $\sigma$ approaches the upper vertex or corner of the equilateral triangle along the midline between  the sides of fixed down spins, the response function decays with power law exponents $47/8$, and for $\epsilon$ the exponent is $5$, with an amplitude that depends on the position of the other operator. It would be interesting to compare these predictions with simulations.   

\vspace{0.5cm}

{\bf ACKNOWLEDGMENT}

\vspace{0.3cm}

I thank T. W. Burkhardt for many useful discussions. 

\vspace{0.5cm}


\appendix

\section{OPERATOR EXPANSION OF THE PRODUCT $\sigma \times \epsilon$ UP TO FOURTH ORDER}  \label{OPE3}

Unlike the Cartesian language used in Eqs. (\ref{next}) for the OPE in low order, for higher order it is advantageous to use the complex notation, see Ref. \cite{BPZ}. Extending the OPE (\ref{next}) by two more orders the result can be written as 
\begin{eqnarray} \label{ope3}
	&&\sigma(z_1,\bar{z}_1) \, \epsilon(z_1+z_{21},\bar{z}_1+\bar{z}_{21}) \, \to \, -{1 \over 2} {1 \over |z_{21}|} \times \Bigl[1+{\cal a} \bigl(z_{21} L_{-1} + \bar{z}_{21} \bar{L}_{-1} \bigr) +\nonumber \\
	&&+{\cal b} \bigl(z_{21}^2 L_{-1}^2 + \bar{z}_{21}^2 \bar{L}_{-1}^2 \bigr) + {\cal c} |z_{21}|^2  L_{-1} \bar{L} _{-1} +  \\
	&&+ \alpha \bigl(z_{21}^3 L_{-3} + \bar{z}_{21}^3 \bar{L} _{-3} \bigr) + \beta \bigl(z_{21}^3 L_{-1}^3 + \bar{z}_{21}^3 \bar{L}_{-1}^3 \bigr) + \gamma \bigl(z_{21}^2 \bar{z}_{21}  L_{-1}^2 \bar{L} _{-1} +  \bar{z}_{21}^2 z_{21} \bar{L}_{-1}^2 L_{-1} \bigr) \Bigr] \times \sigma(z_1,\bar{z}_1)  \nonumber
\end{eqnarray}
where 
\begin{eqnarray} \label{ope3'}
	{\cal a}=4 \, , \quad {\cal b}={8 \over 3} \, , \quad {\cal c}=16
\end{eqnarray}
and
\begin{eqnarray} \label{ope3prime}
	\alpha = - {4 \over 7} \, , \quad \beta = {32 \over 21} \, , \quad \gamma = {32 \over 3} \, . 
\end{eqnarray}
Since $\sigma \times \epsilon $ is odd on reversing all Ising spins the expansion (\ref{ope3}) is in terms of $\sigma$ and its descendants. These arise from it by repeatedly applying the operations \cite{BPZ}
\begin{eqnarray}
\label{def_descendant}
L_{-p} \,\Sigma(z_{1},\bar{z}_{1}) \equiv \int_{\mathcal{C}_{z_{1}}} \frac{\textrm{d}z}{2\pi i} (z-z_{1})^{-p+1} T(z) \, \Sigma(z_{1},\bar{z}_{1}), \nonumber \\
\bar{L}_{-p}\, \Sigma(z_{1},\bar{z}_{1}) \equiv \int_{\mathcal{C}_{\bar{z}_{1}}} \frac{\textrm{d}\bar{z}}{2\pi i} (\bar{z}-\bar{z}_{1})^{-p+1} \bar{T}(\bar{z}) \, \Sigma(z_{1},\bar{z}_{1}) \, ,
\end{eqnarray}
with $p=1,2,3,...$. Here $\mathcal{C}_{z_{1}}$ and ${\mathcal{C}_{\bar{z}_{1}}}$ are closed integration paths enclosing counterclockwise the points $z_{1}$ and $\bar{z}_{1}$, respectively. In particular, $L_{-1} \, \Sigma(z_{1},\bar{z}_{1}) = \partial_{z_1} \, \Sigma(z_{1},\bar{z}_{1})$. Consecutive operations do in general not commute, but follow the Virasoro algebra \cite{BPZ}, since the integration path of the operation to the right is nested inside the integration path of the one to the left. Due to the degeneracy of $\sigma$ on level 2 in the Ising model, $L_{-2} \, \sigma = (4/3) L_{-1}^2 \, \sigma$, and, choosing $L_{-1}^2 \, \sigma$, the operator $L_{-2} \, \sigma$ does not appear in (\ref{ope3}). Moreover, $L_{-2} \, L_{-1} \, \sigma$ does not appear since it can be expressed via the Virasoro algebra \cite{BPZ} in terms of $L_{-1} L_{-2} \, \sigma$ and $L_{-3} \, \sigma$, i.e. in terms of $L_{-1}^3 \, \sigma$ and $L_{-3} \, \sigma$.  . 

In the following the second, third, and fourth order terms in Eq. (\ref{ope3}) will often be addressed as the first, second, and third corrections.

\subsection{Derivation by comparing with the four-point function in the bulk}
 
The operator form in (\ref{ope3}) is consistent with the general expression of the OPE for two primary operators given in Ref. \cite{BPZ}, and the prefactors that are specific for the product $\sigma \times \epsilon$ in the Ising model can be obtained by comparing with the bulk four point function $\langle \sigma(1) \,  \epsilon(2) \, \sigma(3) \, \epsilon(4) \rangle$ taken from Eq. (32) of Ref. \cite{Mattis}. In particular, to obtain the values in (\ref{ope3prime}), we expand the four point function with $z_1 =0$ and $z_{21}=1$ for large $|z_3|, |z_4|$ and find in the order of the third correction the result
\begin{eqnarray} \label{fourpoint}
\langle \sigma(0,0) \,  \epsilon(1,0) \, \sigma(z_3,\bar{z}_{3}) \, \epsilon(z_4,\bar{z}_{4}) \rangle_{\rm bulk} \big|_{\rm 3rd \, corr}= -{1 \over 16} {\Theta} \, {\rm Re} \Bigl[&& -7z_3^{-3}+32z_4^{-3}-16z_3^{-1}z_4^{-2}-4z_3^{-2}z_4^{-1} \nonumber \\
&&+15z_3^{-2} \bar{z_3}^{-1}+64z_4^{-2} \bar{z_4}^{-1} \nonumber \\
&&-48z_3^{-1} \bar{z_4}^{-2}-20z_3^{-2} \bar{z_4}^{-1} \nonumber \\
&&+24|z_3|^{-2}z_4^{-1}-32|z_4|^{-2}z_3^{-1} \, \Bigr]
\end{eqnarray}
where \blue{\cite{higher}}
\begin{eqnarray} \label{fourpointprime}
{\Theta}=\langle \sigma(z_3, \bar{z}_3) \epsilon(z_4, \bar{z}_4) \sigma (0,0) \rangle_{\rm bulk} \equiv -{1 \over 2}|z_3|^{3/4}|z_4|
^{-1}|z_3 -z_4|^{-1} \, .
\end{eqnarray}
Due to the general bulk relation \cite{misprint}
\begin{align}
\label{cf_002}
2 \textrm{Re} \bigl\langle \sigma(z_{3},\bar{z}_{3}) \epsilon(z_{4},\bar{z}_{4}) \times (-1/2) \left( [\alpha L_{-3}+\beta L_{-1}^{3}+\gamma L_{-1}^{2}\bar{L}_{-1}]\sigma(z_{1},\bar{z}_{1}) \right) \bigr\rangle_{\rm bulk} \big\vert_{z_{1}=0} = \Theta \textrm{Re}(\mathcal{T}) \, ,
\end{align}
with
\begin{eqnarray} \nonumber
\label{cf_003}
\mathcal{T} & = & \frac{\alpha}{4} \left( z_{3}^{-3} - 6 z_{4}^{-3} +2z_{3}^{-1}z_{4}^{-2}+2z_{3}^{-2}z_{4}^{-1} \right) + \\ \nonumber
& + & \frac{3 \beta}{512} \left( 65 z_{3}^{-3} - 320 z_{4}^{-3} + 144 z_{3}^{-1}z_{4}^{-2} + 60 z_{3}^{-2}z_{4}^{-1} \right) + \\ \nonumber
& + & \frac{3 \gamma}{512} \left( -15 z_{3}^{-2}\bar{z}_{3}^{-1} - 64 z_{4}^{-2}\bar{z}_{4}^{-1} + 20 z_{3}^{-2}\bar{z}_{4}^{-1} + 48 z_{4}^{-2}\bar{z}_{3}^{-1} - 24 |z_{3}|^{-2} z_{4}^{-1} + 32 |z_{4}|^{-2} z_{3}^{-1} \right) \, , \\
\end{eqnarray}
the OPE (\ref{ope3}) is consistent with the result (\ref{fourpoint}), (\ref{fourpointprime}) of the four point function if $\alpha, \beta,\gamma$ take the values given in Eq. (\ref{ope3prime}). 

\subsection{Checking against the three-point function}

Expanding the three-point function \cite{higher} in complex notation,
\begin{eqnarray} \label{threecomplex}
\langle \sigma (0,0) \epsilon(z_2 = x, \bar{z}_2=x) \sigma(z_3, \bar{z}_3)   \rangle_{\rm bulk} \equiv -{1 \over 2} \, {(z_3 \bar{z}_3)^{3/8} \over |x| (z_3 -x)^{1/2} (\bar{z}_3 -x)^{1/2}} 
 \, ,
\end{eqnarray}
up to order $x^3 /|x|$ yields
\begin{eqnarray} \label{threecomplexprime}
\langle \sigma (0,0) \epsilon(x,x) \sigma(z_3, \bar{z}_3) \rangle_{\rm bulk} \to -{1 \over 2} \, {1 \over |x| |z_3|^{1/4}} &\times& \Bigl\{ 1+ x {\rm Re} z_3^{-1}  + {x^2 \over 4} \bigl( 3{\rm Re}z_3^{-2} + |z_3|^{-2} \bigr) + \nonumber \\
&& \qquad +{x^3 \over 8} \, \bigl(5 {\rm Re} z_3^{-3} + 3 {\rm Re} z_3^{-1} \bar{z} _3^{-2} \bigr) \Bigr\}
\, .
\end{eqnarray}

First we confirm that the term $\propto x^2 / |x|$ in (\ref{threecomplexprime}) is reproduced by multiplying the sum of the $\cal b$ and $\cal c$ terms in (\ref{ope3}) with $\sigma (z_3 , \bar{z}_3)$, taking the bulk average, and finally putting $z_{21}=x$ and $z_1 =0$. Using that $\langle \sigma (z_1, \bar{z}_1) \, \sigma (z_3, \bar{z}_3) \rangle_{\rm bulk} = (z_1 - z_3)^{-1/8} (\bar{z}_1 - \bar{z}_3 )^{-1/8}$ one realizes that the first and second term in the bracket that multiplies $x^2 /4$ in (\ref{threecomplexprime}) follows from the $\cal b$ and $\cal c$ term, respectively.

Next consider the term $\propto  x^3/|x|$, for which the OPE (\ref{ope3}) predicts
\begin{eqnarray} \label{threepredict}
&& \qquad \langle \sigma (0,0) \epsilon(x,x) \sigma(z_3, \bar{z}_3) \rangle_{\rm bulk} \big|_{\rm 3rd \, corr} = -{1 \over 2} \, {x^3 \over |x|} \times {\cal S} \, , \nonumber \\
&&{\cal S} \equiv 2 \textrm{Re} \bigl\langle \sigma(z_{3},\bar{z}_{3}) \left( [\alpha L_{-3}+\beta L_{-1}^{3}+\gamma L_{-1}^{2}\bar{L}_{-1}]\sigma(z_{1},\bar{z}_{1}) \right) \bigr\rangle_{\rm bulk} \big\vert_{z_{1}=0}
\, .
\end{eqnarray}
Using the relationship
\begin{eqnarray} \label{relate3}
{\cal S} =  2 \textrm{Re} \biggl[ \left( \frac{\alpha}{4} + \frac{153}{512} \beta \right) z_{3}^{-3} + \frac{9 \gamma}{512} z_{3}^{-2}\bar{z}_{3}^{-1} \biggr] |z_{3}|^{-1/4}
\, ,
\end{eqnarray}
see Eq. (B.1) in the paper of Ref. \cite{misprint}, together with the prefactors in (\ref{ope3prime}), one verifies that the prediction (\ref{threepredict}) reproduces the term $\propto x^3 / |x|$ in Eq. (\ref{threecomplexprime}).

\subsection{Applying the OPE in the upper half plane with uniform boundary condition +} \label{opuniform} 

A simple check of the OPE in (\ref{ope3}) in the presence of a boundary is provided by averaging it in the upper half plane with a {\it uniform} boundary condition + and comparing the result with the exact result of $\langle \sigma \epsilon \rangle_+$ in Eq. (\ref{+epssig}). While the expansion about $z_1$ in direction $z_{21}=x_{21}$ contains only even powers, the one in direction $z_{21}=i y_{21}$ contains both even and odd powers in $y_{21}$. In the latter case Eq.  (\ref{+epssig}) yields
\begin{eqnarray} \label{uni}
&&\langle \sigma(z_{1},\bar{z}_{1}) \epsilon(z_{1}+iy_{21},\bar{z}_{1}-iy_{21}) \rangle_+ \to -{1 \over 2 |y_{21}|} \times \nonumber \\
&& \qquad \qquad \times \Bigl(1-{1 \over 2} Y + {3 \over 4} Y^2 -{7 \over8} Y^3 + ... \Bigr) \, \langle \sigma(z_1 , \bar{z}_1) \rangle_+ \,   
\end{eqnarray}
where $Y \equiv y_{21} / y_1 $. 

To compare with the OPE, we start with the $Y^2$ term in (\ref{uni}) which on using the expression $\langle \sigma(z_1 , \bar{z}_1) \rangle_+ = \bigl(4i /(z_1 - \bar{z}_1) \bigr)^{1/8}$ is reproduced by the sum of the $\cal b$ and $\cal c$ terms in (\ref{ope3}). Here the contribution of the $\cal b$ term is by a factor 3 smaller than the contribution of the $\cal c$ term.

Now consider the $Y^3$ term in (\ref{uni}) to be reproduced by the contributions from the $\alpha$, $\beta$, and $\gamma$ terms in the OPE (\ref{ope3}). First we evaluate $\langle L_{-3} \sigma(z_1, \bar{z}_1) \rangle_+$. For later use we note the expressions for the more general case of the $+-$ boundary condition where
\begin{eqnarray} \label{L3+-}
\langle L_{-3} \, \sigma(z_1, \bar{z}_1) \rangle_{+-}= \int_{\mathcal{C}_{z_{1}}} \frac{\textrm{d}z}{2\pi i} (z-z_{1})^{-2} \langle T(z) \, \sigma(z_{1},\bar{z}_{1}) \rangle_{+-} 
\end{eqnarray}
with \cite{BX}
\begin{eqnarray} \label{3+-}
&&\langle T(z) \sigma(z_{1},\bar{z}_{1}) \rangle_{+-} = \bigl[(1)+(2)+(3)+(4)+(5)+(6) \bigr] \langle \sigma(z_{1},\bar{z}_{1})\rangle_{+-}^{(\zeta)} \Big|_{\zeta =0}  \, , \nonumber \\
&&(1)= {1/16 \over (z-z_1)^2} \, , \quad  (2) = {1 \over z-z_1} \partial_{z_1} \, , \quad  (3)=  {1/16 \over (z-\bar{z}_1)^2} \, , \quad  (4) = {1 \over z-\bar{z}_1} \partial_{\bar{z}_1} \, , \nonumber \\
&&(5) = \langle T(z) \rangle_{+-}  \, , \quad (6) ={ 1 \over z-\zeta} \partial_{\zeta} \, .
\end{eqnarray}
Here $\zeta$ is the switching point on the boundary. Note that the terms (1) and (2) do not contribute to the integral in (\ref{L3+-}). For the present uniform + boundary the average of $\sigma$ is $\langle \sigma(z_{1},\bar{z}_{1})\rangle_+ = (4i/(z_1-\bar{z}_1))^{1/8}$ and the terms (5) and (6) are absent. Expanding (3) and (4) about $z_1$ to first order in $z-z_1$ yields
\begin{eqnarray} \label{L3+}
\langle L_{-3} \, \sigma(z_1, \bar{z}_1) \rangle_{+}= -{1 \over 4} {1 \over (z_1 - \bar{z}_1)^3} \, \langle \sigma(z_{1},\bar{z}_{1}) \rangle_{+} \, .
\end{eqnarray}
The two remaining third-order terms in (\ref{ope3}) are easily evaluated since $L_{-1} = \partial_{z_1}$ and $\bar{L}_{-1} = \partial_{\bar{z}_1}$ and one finds
\begin{eqnarray} \label{L3+prime}
\langle \Bigl(L_{-3} \, , \, L_{-1}^3 \, , \, L_{-1}^2 \bar{L}_{-1} \Bigr) \, \sigma(z_1, \bar{z}_1) \rangle_{+}= \Bigl(-{1 \over 4} \, ,\, -{153 \over 512} \, , \, {153 \over 512}  \Bigr)\times {i \over 8 y_1^3} \, \langle \sigma(z_1 , \bar{z}_1) \rangle_+
\end{eqnarray}
which inserted in (\ref{ope3}) for $z_{21} = i y_{21}$ yields
\begin{eqnarray} \label{L3+primeprime}
\langle \sigma \epsilon \rangle_+ \Big|_{\propto Y^3} = -{1 \over 2 |y_{21}|} \,  \, {y_{21}^3 \over 8 y_1^3}  \times 2 \Bigl[-{1 \over 4} \alpha -{153 \over 512} (\beta + \gamma) \Bigr] \, \langle \sigma(z_1 , \bar{z}_1) \rangle_+.
\end{eqnarray}
Substituting the values (\ref{ope3prime}), the square bracket equals $-7/2$ and the rhs of (\ref{L3+prime}) indeed reproduces the $Y^3$ term in Eq. (\ref{uni}).

\subsection{Applying the OPE in the upper half plane with mixed boundary condition $+-$} \label{OPE+-}

\subsubsection{Case $x_1=0$}

Here we show that the OPE (\ref{ope3}) reproduces not only the leading but also the next-to-leading behaviors of $\langle \sigma (x_1=0, y_1=y_0) \, \epsilon (x_2, y_2=y_0) \rangle_{+-}$ near $x_2 = 0$ which are displayed in Eq. (\ref{half+-limits}). In the complex notation of the OPE (\ref{ope3}) we have 
\begin{eqnarray} \label{x1eq0}
z_1 = i y_0 \, , \, z_2 = iy_0+x_2 \quad {\rm so \, that} \quad z_{21}=x_2
\end{eqnarray}
and we shall need the expressions \cite{BX}
\begin{eqnarray} \label{sig+-}
\langle \sigma (z_1, \bar{z}_1) \rangle_{+-} = - \Bigl({4i\over z_1-\bar{z}_1}\Bigr)^{1/8} \times {1 \over (z_1 \bar{z}_1)^{1/2}} \times {z_1 + \bar{z}_1 \over 2} \, , \nonumber \\
\partial_{\zeta} \langle \sigma (z_1, \bar{z}_1) \rangle_{+-}^{(\zeta)}\big|_{\zeta =0} = 2^{1/8} \Bigl({2i\over z_1-\bar{z}_1}\Bigr)^{1/8 -2} \times {1 \over (z_1 \bar{z}_1)^{3/2}} \, .
\end{eqnarray}
Calculating $\langle L_{-3} \, \sigma \rangle_{+-}$ along Eqs. (\ref{L3+-}) and (\ref{3+-}) the terms (3) and (5) do not contribute since they are proportional to $\langle \sigma \rangle_{+-}$ which vanishes for $x_1 =0$ so that only the terms (4) and (6) survive. Due to $\bigl(\partial_{\bar{z}_1}, \partial_{\zeta}\bigr) \langle \sigma (z_1, \bar{z}_1) \rangle_{+-} = \bigl(-1/2, 1\bigr) (2/y_0)^{1/8} y_0^{-1}$ for our $z_1 = i y_0$, expanding the prefactors in (4) and (6) to first order in $z-z_1$ yields
\begin{eqnarray} \label{L3+-prime}
\langle L_{-3} \, \sigma(z_1, \bar{z}_1) \rangle_{+-} \big|_{z_1 = i y_0} = {7 \over 8} \, {1 \over y_0^3} \, \Bigl( {2 \over y_0} \Bigr)^{1/8} \, .
\end{eqnarray}
For the averages of the other two descendants of $\sigma$ one finds
\begin{eqnarray} \label{otherL3+-prime}
\langle \bigl(L_{-1}^3 \, , \, L_{-1}^2 \bar{L}_{-1}\bigr)  \, \sigma(z_1, \bar{z}_1) \rangle_{+-} \big|_{z_1 = i y_0} = {3 \over 512} \, \bigl(217 \, , \, 13  \bigr) {1 \over y_0^3} \, \Bigl( {2 \over y_0} \Bigr)^{1/8} \, .
\end{eqnarray}
Here one uses $L_{-1} = \partial_{z_1}$, $\bar{L}_{-1} = \partial_{\bar{z}_1}$, and the differentiations are simplified since for a nonvanishing result for $z_1 = i y_0$ the last factor in the upper equation (\ref{sig+-}) must always be differentiated. Substituting (\ref{L3+-prime}) and (\ref{otherL3+-prime}) in (\ref{ope3}) yields
\begin{eqnarray} \label{nextto+-}
\langle \sigma \epsilon \rangle_{+-} \big|_{z_1 = i y_0, \, z_2 = i y_0 +x_2} \big|_{\rm 3rd \, corr} = - {1 \over 2} {1 \over |x_2|} \Bigl( {x_2 \over y_0} \Bigr)^3 \, \Bigl( {2 \over y_0} \Bigr)^{1/8} \times 2 \Bigl[{7 \over 8} \alpha  +{3 \cdot 217 \over 512}  \beta  +  {39 \over 512} \gamma \Bigr] .
\end{eqnarray}
Inserting (\ref{ope3prime}) the square bracket takes the value $9/4$ and(\ref{nextto+-}) reproduces the next-to-leading order term for small $|x_2|$ in (\ref{half+-limits}).

\subsubsection{Case $x_2 =0$}

Here we have
\begin{eqnarray} \label{x2eq0}
z_1 = i y_0 + x_1 \, , \, z_2 = iy_0 \quad {\rm so \, that} \quad z_{21}=-x_1 \, .
\end{eqnarray}
When correspondingly inserting $z_1 = i y_0 + x_1, \, z_{21}=-x_1$ in the OPE (\ref{ope3}) and taking the average $\langle \, \rangle_{+-}$ the rhs displays an $x_1$ dependence proportional to $(1/|x_1|) \, [x_1, \, x_1^3, \, x_1^5, \, {\rm etc}]$, consistent with the exact result of the two-point function $\langle \sigma \, \epsilon \rangle_{+-}$ in Eq. (\ref{half+-prime}). The reason is that even powers of $z_{21}= \bar{z}_{21}=-x_1$ are accompanied by averages of ${\sigma}$-descendants that are odd in $x_1$ and vice versa. The leading order contribution to $\langle \sigma \, \epsilon \rangle_{+-}$ comes from the first two terms in the square bracket in (\ref{ope3}) which yields
\begin{eqnarray} \label{x2eq0lead} 
\langle \sigma (i y_0 + x_1 , -i y_0 + x_1 ) \, \epsilon (i y_0 , -iy_0) \rangle_{+-} \to -{1 \over 2|x_1|} \, \Bigl( {2 \over y_0} \Bigr)^{1/8} \, {x_1 \over y_0} \, \bigl( -1 + {\cal a} \bigr) 
\end{eqnarray}
with ${\cal a}=4$ from (\ref{ope3'}). This reproduces the leading behavior of the two-point function given in Eq. (\ref{half+-prime}).

\subsection{OPE in the upper half plane with a $-+-$ boundary} \label{OPE-+-}

Now we apply the OPE (\ref{ope3}) about the point $g_1=0, \,  j_1 =1$ in the upper half $h$ plane with a $-+-$ boundary where $\langle \sigma \rangle_{-+-}$ vanishes and show that it reproduces the expansion (\ref{h1equiprime}). Here $h_1 =i$ and $h_{21} = g_2 +i (j_2 -1)$. 

The contribution to $\langle \sigma \, \epsilon \rangle_{-+-}$ in (\ref{h1equiprime}) from the second term in its curly bracket follows from the ${\cal b}$ and ${\cal c}$ terms in (\ref{ope3}). For the evaluation it helps to rewrite (\ref{1sig-+-}) in the form
\begin{eqnarray} \label{si-+-} 
\langle \sigma (h_1, \bar{h}_1 \rangle_{-+-} = (1-|h_1|^2) \times {\cal D} \, , \quad {\cal D} \equiv  \Bigl( {4i \over h_1-\bar{h}_1} \Bigr)^{1/8} \; (1-h_1^2)^{-1/2} \, (1-\bar{h}_1^2)^{-1/2}
\end{eqnarray}
and to observe that \cite{L1-+-}
\begin{eqnarray} \label{derivsig-+-} 
&&\langle L_{-1}^2 \sigma \rangle_{-+-} \big|_{h_1=i} \equiv  \partial_{h_1}^2 \langle \sigma \rangle_{-+-} \big|_{h_1=i} = - 2 \bar{h}_1 \partial_{h_1} {\cal D} \big|_{h_1=i} = - 2^{1/8} \, 9/16 \, , \nonumber \\
&&\langle L_{-1} \bar{L}_{-1} \sigma \rangle_{-+-} \big|_{h_1=i} \equiv \partial_{h_1} \partial_{\bar {h}_1} \langle \sigma \rangle_{-+-} \big|_{h_1=i} = \nonumber \\
&& \qquad  \qquad \qquad = - \bigl[ 1+(h_1 \partial_{h_1} + \bar {h}_1 \partial_{\bar {h}_1}) \bigr] {\cal D} \big|_{h_1=i} = 2^{1/8} /16 \, .
\end{eqnarray}
Together with the above form of $h_{21}$ and the coefficients (\ref{ope3'}) the OPE (\ref{ope3}) in the $h$ plane then leads to the desired contribution in (\ref{h1equiprime}). 

The contribution to $\langle \sigma \, \epsilon \rangle_{-+-}$ in (\ref{h1equiprime}) from the third term in its curly bracket follows from the $\alpha$, $\beta$, and $\gamma$ terms in the OPE (\ref{ope3}) and is reproduced by using the results
\begin{eqnarray} \label{OPE3-+-} 
\langle \bigl( \alpha L_{-3} , \, \beta L_{-1}^3 , \, \gamma L_{-1}^2 \bar{L}_{-1} \bigr) \, \sigma (h_1 , \, \bar{h}_1 ) \rangle_{-+-} \big|_{h_1 = i} = - \,  2^{1/8} {i \over 16 \times 7} \, (-40, \, 89, \, 49) \, .
\end{eqnarray}

Finally consider the OPE about the point $g_2 =0, \, j_2 =1$, i.e. $h_2 =i$. Here (\ref{ope3}) yields 
\begin{eqnarray} \label{h2OPEshort}
\langle \sigma (g_1 =0 , j_1) \, \epsilon (g_2 =0 , j_2 =1) \rangle_{-+-}  &\to&  - {2^{(1/8)+1} \over 4|j_1 -1|} \, \Bigl[ (j_1 -1) \, (-1+ {\cal a}) + \nonumber \\
&&+ (j_1 -1)^2 \, \Bigl( {5 \over 8} - {5 \over 4} {\cal a} + {9 \over 8} {\cal b} + {1 \over 16} {\cal c}  \Bigr) \Bigr]
\end{eqnarray}
which on inserting the values given in (\ref{ope3'}) reproduces (\ref{h2equishort}).

\subsection{OPE in the triangle} \label{OPEtri}

Here we derive the result (\ref{shorttri1at0}) within the OPE, showing how it follows from the ${\cal a} \, , \, {\cal b}$, and ${\cal c}$-terms. For the present purpose let us write Eq. (\ref{ope3}) in the form
\begin{eqnarray} \label{optri} 
\langle \sigma (z_1 =iy_0,\bar{z}_1) \, \epsilon(z_2 =z_1 +z_{21} , \bar{z}_2) \rangle_{\rm triangle}^{(\rm cum)} \, \to \, \Bigl({ Y_{\rm C} \over y_{\rm C}} \Bigr)^{(1/8)+1} \Bigl( -{1\over 2 |Z_{21}|} \Bigr) \times \nonumber \\
\times \Bigl[ {\cal a} (Z_{21} A + {\rm cc}) +{\cal b} (Z_{21}^2 B + {\rm cc}) + {\cal c} |Z_{21}|^2  C \Bigr]
\end{eqnarray}
where 
\begin{eqnarray} \label{optriprime} 
\big(A,B,C \bigr) = \big(\partial_{Z_1} \, , \, \partial_{Z_1}^2 \, , \, \partial_{Z_1} \partial_{\bar{Z}_1} \bigr) \times \bigl( S(Z_1) S(\bar{Z}_1 \bigr)^{1/16} \langle \sigma (h_1 , \bar{h}_1) \rangle_{-+-} \, \Big|_{Z_1 = i Y_0} .
\end{eqnarray}
Expressing everything in terms of $h$ via $\partial_{Z_1} = S(Z_1) \partial_{h_1} $ one finds 
\begin{eqnarray} \label{optriA} 
A=|S|^{1/8} S \, \bigl( \partial_{h_1} \langle \sigma \rangle_{-+-} \bigr) \big|_{h_1 = i} =S_0^{(1/8)+1} \, 2^{(1/8)-1} \, i \, , \nonumber \\
B=\bar{S}^{1/16} \Bigl[ {9 \over 8} S^{(1/16)+1} \bigl(\partial_{h_1} S \bigr) \, \bigl( \partial_{h_1} \langle \sigma \rangle_{-+-} \bigr)\, + \, S^{(1/16)+2} \, \bigl( \partial_{h_1}^2 \langle \sigma \rangle_{-+-} \bigr) \Bigr] \Big|_{h_1 =i} \, , = \nonumber \\ 
=- \, S_0^{(1/8)+2} \, 2^{(1/8) -1} \, {3 \over 8} \, \nonumber \\
C=\Bigl[{1 \over 16} \bigl(|S|^{1/8} \bar{S} \, (\partial_{h_1} S) \, ( \partial_{\bar{h}_1} \langle \sigma \rangle_{-+-}) + {\rm cc} \bigr) + |S|^{(1/8) +2} \, (\partial_{h_1} \partial_{\bar{h}_1}\langle \sigma \rangle_{-+-}) \Bigr] \Big|_{h_1 =i} = \nonumber \\
= S_0^{(1/8)+2} \, 2^{(1/8)-1} \, {1 \over 24} \, .
\end{eqnarray}
Here in the last steps we use the expressions (\ref{resc}) for $S$ and \cite{L1-+-}, (\ref{si-+-}), (\ref{derivsig-+-}) for the $h$ derivatives of $\langle \sigma \rangle_{-+-}$. Substituting $A,B,C$ from (\ref{optriA}) and ${\cal a,b,c}$ from (\ref{ope3'}) in (\ref{optri}) and putting $Z_{21} = i (Y-Y_0)$ one arrives at the small-distance expression (\ref{shorttri1at0}).
\subsection{OPE in the square} \label{OPEsq}
Here we address the leading short distance behavior of the  two-point functions $\langle \sigma \epsilon \rangle_{\rm SQ}$ in the square of Sec. \ref{-+-+square} with one of the operators located in the square's center. Their forms are given in Eqs. (\ref{originsigmaepsilon})-(\ref{originepsilon}) and are derived from the order parameter profile (\ref{sigSQ}) using the OPE (\ref{ope3}). While for $\sigma$ in the center only the ${\cal b}$-term contributes, the results for $\epsilon $ in the center follow from the three terms 1, $\propto {\cal a}$, and $\propto {\cal b}$ in the square bracket of Eq. (\ref{ope3}) yielding
\begin{eqnarray} \label{epsincenter} 
\langle \sigma (z_1, \bar{z}_1) \, \epsilon (0,0) \rangle_{\rm SQ} \big/ \langle \sigma (z_1, \bar{z}_1) \rangle_{\rm SQ} \to - {1 \over 2 |z_1| } \, \bigl[1 -2 {\cal a} +2 {\cal b}\bigr] \, .
\end{eqnarray}
This leads to the second equation in (\ref{originsigmaepsilon}).


\section{CUMULANTS WITH THE STRESS TENSOR AND NEAR-BOUNDARY BEHAVIOR} \label{stresscum}

Cumulants with the stress tensor follow from the conformal Ward identity. For the half plane with $+-$ boundary condition, see Eq. (2.4) in Ref. \cite{BX} which is reproduced in the present Eq. (\ref{3+-}). We are interested in the cumulants $\langle T \, \phi \rangle^{(\rm cum)}$ with $\phi = \sigma$ or $\epsilon$. Besides their importance for the OPE as described in Appendix \ref{OPE3}, via the BOE \cite{BOE} they describe the near boundary behavior of the response function $\langle \sigma \epsilon \rangle_{+-}$.

The profiles $\langle \phi \rangle$ and their derivatives with respect to the switch point that are needed for the Ward identity can be taken from Eq. (4.1) in Ref. \cite{BX} and from Eqs. (3.28) in Ref. \cite{BE21}, respectively. The results can be written as 
\begin{eqnarray} \label{Tsigma+-} 
\langle T(z) \sigma (x_1 , y_1) \rangle_{+-}^{\rm (cum)} &=& - \Bigl( {4i \over z_1 - \bar{z}_1} \Bigr)^{1/8} \, {1 \over 16} \times \Biggl[ \Bigl( {1 \over (z-z_1)^2} + {1 \over (z-\bar{z}_1)^2} \Bigr) {z_1 +\bar{z}_1 \over 2 |z_1|} +\nonumber \\
&& \qquad \qquad \qquad  + {1 \over z-z_1} P + {1 \over z-\bar{z}_1} \bar{P} + {4 \over z} {(z_1 -\bar{z}_1)^2 \over |z_1|^3} \Biggr]
\end{eqnarray}
where 
\begin{eqnarray} \label{Tsigma+-pri} 
P \equiv {1 \over z_1-\bar{z}_1} \, { 3z_1^2 -9|z_1|^2 + 4 \bar{z}_1^2 \over z_1 |z_1|}
\end{eqnarray}
and
\begin{eqnarray} \label{Tepsilon+-} 
\langle T(z) \epsilon (x_2 , y_2) \rangle_{+-}^{\rm (cum)} =  \Biggl\{ \Bigl( {1 \over (z-z_2)^2} + {1 \over (z-\bar{z}_2)^2} \Bigr) {1 \over 2 i} \Bigl[ {1 \over z_2- \bar{z_2}} +{1 \over \bar{z}_2} -{1 \over z_2} \Bigr] + \nonumber \\
+ {1 \over z-z_2} Q + {1 \over z-\bar{z}_2} \bar{Q} -{1 \over z i} \Bigl( {1 \over z_2^2} - {1 \over \bar{z}_2^2 }\Bigr)  \Biggr\}
\end{eqnarray}
where
\begin{eqnarray} \label{Tepsilon+-pri} 
Q \equiv {1 \over i}\Bigl( {1 \over z_2^2} - {1 \over (z_2 -\bar{z}_2)^2} \Bigr) \, .
\end{eqnarray}
We have checked the relations,
\begin{eqnarray} \label{epsby} 
\langle \sigma (x_1, y_1) \, \epsilon (x_2, y_2) \rangle_{+-}^{(\rm cum)} \to 4 \, y_2  \langle \sigma (x_1, y_1) \, T(x_2) \rangle_{+-}^{(\rm cum)}\, , \quad  y_2 \to 0 
\end{eqnarray}
and
\begin{eqnarray} \label{sigby} 
\langle \sigma (x_1, y_1) \, \epsilon (x_2, y_2) \rangle_{+-}^{(\rm cum)} \to (-2^{1/8}, 2^{1/8}) \, y_1^{2-(1/8)}  \langle T(x_1) \epsilon (x_2, y_2) \,\rangle_{+-}^{(\rm cum)}\, , \quad  y_1 \to 0  
\end{eqnarray}
for $(x_1<0,x_1>0)$, that are implied by the BOE explained in Ref. \cite{BOE}.

As a simple example consider the special case of (\ref{Tsigma+-}),
\begin{eqnarray} \label{Tsigma}
\langle T(z) \, \sigma (x_1 =0, y_1 =y_0) \rangle_{+-} = \Bigl( {2 \over y_0} \Bigr)^{1/8} \,    {y_0 \over z(z^2 +y_0^2)}
\end{eqnarray}
and its relation to the response function
\begin{eqnarray} \label{byse+-}
\langle \sigma (x_1 =0 , y_1 =y_0) \, \epsilon (x_2 , y_2) \rangle_{+-} \equiv \Bigl( {2 \over y_0} \Bigr)^{1/8} \, {4 y_0 x_2 y_2 \over |z_2|^2 \, |z_2^2 +y_0^2|} \, .
\end{eqnarray}
Eq. (\ref{byse+-}) follows from Eq. (4.3) in Ref. \cite{BX} or from (\ref{se+-cum}), (\ref{seA}) for $z_1 =i y_0$ for which $A=0$ and $B+C= -8 y_0^2 y_2^2 x_2$. For $y_2 \to 0$ the rhs of (\ref{byse+-}) tends to $(2/y_0)^{1/8} 4 y_0 y_2 / [x_2 (x_2^2 +y_0^2)]$ which on using (\ref{Tsigma}) is indeed reproduced by the BOE predicting $4 y_2 \langle T(x_2) \sigma (x_1 =0, y_1 =y_0 \rangle_{+-}$ for the lhs of (\ref{byse+-}), see Ref. \cite{BOE}.

The relations (\ref{epsby}) and (\ref{sigby}) imply in particular that the cumulants (\ref{Tsigma+-}) and (\ref{Tepsilon+-}) must be real if $z \to x$ becomes real. Moreover for $z=x$, $\langle T \sigma \rangle_{+-}^{(\rm cum)}$ in (\ref{Tsigma+-}) must be odd in $x$ when $x_1 =0$ and $\langle T \epsilon \rangle_{+-}^{(\rm cum)}$ in  (\ref{Tepsilon+-}) be even in $x$ when $x_2=0$ as it follows from the antisymmetry of $\langle \sigma \epsilon \rangle_{+-}^{(\rm cum)}$ given in (\ref{+-symmetry}). 

Next we discuss the cumulants $\langle T(h)\,  \phi (g_1, j_1) \rangle_{-+-}^{(\rm cum)}$ in the $-+-$ plane. By means of the Moebius mapping (\ref{Möb}) and the usual transformation formula one finds, e.g., in the case of $\phi = \sigma$
\begin{eqnarray} \label{Tsig}
\langle T(h) \, \sigma (g_1 , j_1) \rangle_{-+-}^{(\rm cum)} = -\Bigl( {2 \over j_1} \Bigr)^{1/8} \, {1 \over (h+1)^4}  \, {1 \over 4} \times B(h; \, g_1 , \, j_1) \, .
\end{eqnarray}
Here $B$ is the square bracket in Eq. (\ref{Tsigma+-}) with $z$ and $z_1$ expressed in terms of $h$ and $g_1 , \, j_1$ by replacing $z \to z(h)$ and $z_1 \to z_1 (h_1 = g_1 +i j_1)$ via the Moebius mapping. In the following we consider the case with both $T$ and $\phi$ located on the imaginary axis \cite{im-+-} in which $\langle T\,  \phi \rangle_{-+-}^{(\rm cum)}$ must be {\it real} by mirror symmetry.

Since $z_1 - \bar{z}_1 \to 4ij_1 / (1+j_1^2)$ and $|z_1| \to 1$ in this case, the quantity $P$ in (\ref{Tsigma+-pri}) becomes
\begin{eqnarray} \label{Pj1}
P\, \to \, {1 \over 2ij_1 (1-ij_1)^2} \times \bigl[1+30 j_1^2 +j_1^4 -2ij_1 (1-j_1^2)  \bigr] 
\end{eqnarray}
and since
\begin{eqnarray} \label{zz1}
{1 \over z-z_1} \to {-jj_1 +i(j+j_1) +1 \over 2i(j-j_1)}  \, , && \; {1 \over z-\bar{z}_1} \to {jj_1 +i(j-j_1) +1 \over 2i(j+j_1)} \, ,   \nonumber \\
z_1 + \bar{z}_1 &\to& -2 \, {1-j_1^2 \over 1+j_1^2}
\end{eqnarray}
one finds the explicit form
\begin{eqnarray} \label{sT}
\langle \sigma (g_1 =0 , j_1) \,  T(h=ij)  \rangle_{-+-}^{(\rm cum)} \, = \, \Bigl( {2 \over j_1} \Bigr)^{1/8} \, {j_1^2 \over 4} \, { j^2 (j_1^2 +15) -15 j_1^2 - 1  \over (j^2 -j_1^2)^2 \, (j^2 +1) \, (j_1^2 +1)}	\, . 
\end{eqnarray}
For $j_1 =1$ Eq. (\ref{sT}) is consistent with Eq. (\ref{h1iT2}). 

Likewise one finds from (\ref{Tepsilon+-}) and (\ref{Tepsilon+-pri}) with 
\begin{eqnarray} \label{Qj2}
Q\, \to \, -i (j_2^2 +1)^2 \, \Bigl[ {1 \over (j_2^2 -1+2ij_2)^2} +{1 \over 16 \, j_2^2} \Bigr] 
\end{eqnarray}
that
\begin{eqnarray} \label{Te}
\langle \, T(h=ij) \, \epsilon (g_2 =0 , j_2) \, \rangle_{-+-}^{(\rm cum)} \, = \, j_2 \, { j^2  (j_2^4 + 2j_2^2 - 15) - 15 j_2^4 + 2 j_2^2 + 1  \over (j^2 -j_2^2)^2 \, (j^2 +1) \, (j_2^2 +1)^2} \, .	 
\end{eqnarray}
%


\section{MAPPING AN EQUILATERAL TRIANGLE TO THE UPPER HALF PLANE} \label{trihalf}

The conformal transformation \cite{Kober}
\begin{eqnarray} \label{trhalf} 
h (Z) = 2 \times 3^{3/4} \, {{\rm sn} (Z , k) \, {\rm dn} (Z , k) \over \bigl[ 1+ {\rm cn} (Z , k) \bigr]^2 } \, , \quad k= 2^{-3/2} (1+\sqrt{3}) \nonumber \\
Z={Y_{\rm C} \over y_{\rm C}} \times z \, , \, Y_{\rm C} \equiv 2 {\bf K}(k') \, , \; k' \equiv \sqrt{1-k^2} = {1 \over 2} \, \bigl(2-\sqrt{3}\bigr)^{1/2}
\end{eqnarray}
maps the interior of the equilateral triangle of side length ${\cal W}$ and height $  y_{\rm C}=(\sqrt{3} / 2) {\cal W}$ in the $z=x+iy$ plane with corners at 
\begin{eqnarray} \label{trhalfprime} 
z = z_{\rm A} = - {\cal W}/2, \, z_{\rm B} = {\cal W} /2, \, z_{\rm C} = i y_{\rm C} = i (\sqrt{3} /2) {\cal W} 
\end{eqnarray}
to the upper half $h$ plane where the images of the corners are at
\begin{eqnarray} \label{trhalfprimeprime} 
h=(h_{\rm A}, \, h_{\rm B}, \, h_{\rm C}) \, = \, (-1, \, 1, \, \infty) \, .
\end{eqnarray}
The mapping of the corner $z_{\rm C}$ to $h=\infty$ arises from ${\rm cn} \bigl(2i{\bf K}(\sqrt{1-k^2}), k \bigr) = -1$, a relation that applies for arbitrary $k$. The particular value of $k$ given in Eq. (\ref{trhalf}) takes care that the images of the left and right boundary sides of the triangle are located on the real axis of the half plane. Due to the prefactor $2 \times 3^{3/4}$ they are given by the half lines $- \infty < h < -1$ and $1 < h < + \infty$, respectively, while the remaining interval $-1 < h < 1$ is the image of the triangle's base side. The inverse transformation mapping the upper half $h$ plane to the triangle in the $Z$ plane has the Schwarz-Christoffel form  
\begin{eqnarray} \label{htoZ} 
Z(h) = 2 \times 3^{-3/4} \, \int\limits_{0}^{h} \, dh' \, (1-h'^2)^{-2/3}  \, 
\end{eqnarray}
with a prefactor that follows from comparing (\ref{htoZ}) and (\ref{trhalf}) for $h \to 0$ and $Z \to 0$, respectively.

Note the rescaling factor 
\begin{eqnarray} \label{resc} 
S(Z) \equiv dh/dZ &=& - 2 \times 3^{3/4} \times {2k^2 {\rm sn}^2 + {\rm cn} -2 \over (1+{\rm cn})^2} \,  \nonumber \\
= (dZ/dh)^{-1} &=& {1 \over 2} 3^{3/4} \bigl[ 1-h^2 (Z) \bigr]^{2/3} \, , \quad Z=X+iY \, 
\end{eqnarray}
of the transformation (\ref{trhalf}). Here the argument of ${\rm sn}$ and ${\rm cn}$ is $ \bigl( Z,k \bigr) $.

In the following we concentrate on the vertical midline of the triangle corresponding to $Z=i \times Y$ which is mapped to the vertical midline $h=i \times j$ in the upper half $h$ plane via
\begin{eqnarray} \label{trafomid} 
j (Y) = 2 \times 3^{3/4} \times {{\rm sn} (Y , k') \, {\rm dn} (Y , k') \over \bigl[ 1+ {\rm cn} (Y , k') \bigr]^2 } \, 
\end{eqnarray}
with the inverse transformation
\begin{eqnarray} \label{invtrafomid} 
Y(j) = 2 \times 3^{-3/4} \, \int\limits_{0}^{j} \, dj' \, (1+j'^2)^{-2/3} 
\end{eqnarray}
that follow from Eqs. (\ref{trhalf}) and (\ref{htoZ}), respectively \cite{consistent}. We note the values 
\begin{eqnarray} \label{twomid} 
Y(j= \infty) \equiv Y_{\rm C} \equiv 2 {\bf K}(k') = 3.196284004 \, , \,  Y(j=1) \equiv Y_0= 0.7432642107
\end{eqnarray}
and their ratio $Y_0 / Y_{\rm C} = y_0 / y_{\rm C} = 0.2325401027$. Here $Z_0 =i Y_0$ is the position on the midline where $\langle \sigma \rangle_{\rm triangle}$ in (\ref{sigtri}) vanishes since it is mapped to $h=i$ where $j=1$.

The corresponding rescaling factor reads
\begin{eqnarray} \label{rescmid'} 
S(iY) = 2 \times 3^{3/4} \times  {2k'^2 \, {\rm cn}^2 - {\rm cn} + 2(1-k'^2) \over (1+{\rm cn})^2} = {1 \over 2} \, 3^{3/4} \, \bigl[1+j^2 (Y) \bigr]^{2/3}
\end{eqnarray}
where the argument of ${\rm cn}$ is given by $ \bigl( Y,k' \bigr) $. This follows from Eq. (\ref{resc}).

For $j$ close to 1, i.e. $Y$ close to $Y_0$, Eqs. (\ref{invtrafomid}) and (\ref{rescmid'}) imply up to first order
\begin{eqnarray} \label{closeto} 
j(Y)-1 \to S_0 \, (Y-Y_0) \, , \quad S(iY) \to S_0 \, \Bigl[ 1+ {2 \over 3} S_0 \, (Y-Y_0) \Bigr]
\end{eqnarray}
where
\begin{eqnarray} \label{closetoprime} 
S_0 \equiv S(iY_0) \, = \, 2^{(2/3)-1} \, 3^{3/4} \, .
\end{eqnarray}

Near the base line and near the upper corner $Y \to Y_{\rm C} - \delta$, Eqs. (\ref{trafomid}) and (\ref{rescmid'}) yield 
\begin{eqnarray} \label{jandRnearbase} 
j(Y) \to 3^{3/4} \, {Y \over 2}, \; S(iY) \to 3^{3/4} \, {1 \over 2}    
\end{eqnarray}
and 
\begin{eqnarray} \label{jandRnearC} 
j(Y) \to 3^{3/4} \, ({\delta}/2)^{-3}, \; S(iY) \to  3^{3/4} \, {1 \over 2} \times j(Y)^{4/3} \equiv   3^{3/4} \, {3 \over 2} \; ({\delta}/2)^{-4} \, ,   
\end{eqnarray}
respectively. Written in terms of the height $y_{\rm C}$ of the triangle, Eq. (\ref{trhalf}) yields
\begin{eqnarray} \label{triheight} 
Y ={y \over y_{\rm C}} \, Y_{\rm C} \, , \quad \delta = {y_{\rm C} - y \over y_{\rm C}} \, Y_{\rm C} \, \, .   
\end{eqnarray}
%


\section{Order parameter profile in the square} \label{SQU}

Here we evaluate the order-parameter profile in the ${\cal W} \times {\cal W}$ square adressed in Sec. \ref{-+-+square} which we need to calculate via the OPE (\ref{ope3}) the corresponding two-point function $\langle \sigma \, \epsilon \rangle_{\rm SQ}$ near the center.  The profile follows by conformal transformations like in Eqs. (B7)-(B13) in Ref. \cite{E23} from that in the upper half $H=G+iJ$ plane. For the present boundary conditions in the square the real axis $J=0$ is to be endowed with boundary conditions $-+-+-$ for
\begin{eqnarray} \label{bc} 
- \infty < G < -C, \, -C < G < -c, \, -c < G < c, \, c < G < C, \, C < G < + \infty \, ,
\end{eqnarray}
respectively, where
\begin{eqnarray} \label{bcprime} 
C=\sqrt{2}+1, \, c=\sqrt{2}-1 \, .
\end{eqnarray}
For completeness we note the expression 
\begin{eqnarray} \label{completesig} 
\langle \sigma (H, \bar{H}) \rangle &=& - \Bigl( {4i \over H-\bar{H}} \Bigr)^{1/8} \, {1 \over C^4 + 14 C^2 c^2 +c^4} \times {1 \over |H^4 - (C^2 + c^2) H^2 +C^2 c^2|} \times \nonumber \\
&&\times \bigl\{8C^2 c^2 \big[ 2 (H \bar{H})^2 -(C^2 +c^2)(H^2 + \bar{H}^2) + 2 C^2 c^2 \big] + \nonumber \\ %
&& \qquad +(C^2 - c^2)^2 \bigl[ (H \bar{H})^2 - (C^2 +c^2) H \bar{H} +C^2 c^2 \bigr] \bigr\} 
\end{eqnarray}
that applies to {\it arbitrary} $0<c<C$ which is of interest for calculating the order parameter profile inside a rectangle of arbitrary aspect ratio. One easily checks that on approaching the boundary $H=G$, the rhs of (\ref{completesig}) turns to the expected behavior $-(2/J)^{1/8} \, {\rm sign}(G^2 - C^2) \, {\rm sign}(G^2 - c^2)$. Eq. (\ref{completesig}) follows from Eqs. (17) and (18a) in Ref. \cite{BG}. For the special values in (\ref{bcprime}) it follows that $C^2 c^2 =1, \, C^2 +c^2 = 6, \, (C^2 -c^2)^2 = 32, C^4 +c^4 =34$, and $C^4 +14 C^2 c^2 +c^4 = 48$ and (\ref{completesig}) becomes
\begin{eqnarray} \label{hpsquaresig} 
\langle \sigma (H, \bar{H}) \rangle &=& - \Bigl( {4i \over H-\bar{H}} \Bigr)^{1/8} \, {(H \bar{H})^2 -  H^2 - \bar{H}^2 -4 H \bar{H} +1 \over |H^4 -6 H^2 +1|} \, .
\end{eqnarray}
The rhs of Eq. (\ref{hpsquaresig}) vanishes for $H=i$ which is the preimage of the center of the square. The Möbius transformation
\begin{eqnarray} \label{hpcircle} 
H(w) = i \,  {1+w\over 1-w} \, , \, {dH \over dw} = {2i \over (1-w)^2}
\end{eqnarray}
relates the upper half $H$ plane to the interior of the unit circle in the $w=u+iv$ plane. Since $H\bigl(w \equiv\exp (i \alpha) \bigr)= - {\rm cot} (\alpha /2) $, the points $\exp (i \alpha)$ on the periphery of the circle with $\alpha$-values $\pi /4, \, 3 \pi /4, \, 5 \pi /4, \, 7 \pi /4$ map to the switch points $-C, \, -c, \, c, \, C$ on the real axis $H=G$ in Eq. (\ref{bc}) with (\ref{bcprime}) taken into account. Moreover for values $0, \,\pi/2, \pi, 3 \pi/2$ the points map to $- \infty, \, -1, \, 0, \, 1$, respectively. The center $w=0$ of the circle maps to $H=i$. Finally the Schwarz-Christoffel transformation 
\begin{eqnarray} \label{ciclesquare} 
z(w)  = \Lambda \, \int\limits_{0}^{w} \, {d \omega \over \sqrt{1+ \omega^4}}
\end{eqnarray}
maps the interior of the circle to the interior of our ${\cal W} \times {\cal W}$ square in the $z$ plane. Here $\Lambda$ is from below (\ref{originepsilon}). Since the abovementioned points with $\pi /4, \, 3 \pi /4, \, 5 \pi /4, \, 7 \pi /4$ on the boundary of the circle are mapped \cite{CtoS} to the corners $z=({\cal W}/2) \, \bigl[1+i, \, -1+i,\, -1-i,\, 1-i \bigr]$ and those with $0, \,\pi/2, \pi, 3 \pi/2$ to the side-midpoints $z=({\cal W}/2) \, \bigl[1, \, i,\, -1,\, -i \bigr]$ of the square, it has the desired boundary conditions.

To derive the one-point function in (\ref{sigSQ}) we must understand only the prefactor since the dependence on $x_1 , y_1$ is dictated by the symmetries (\ref{mirrormidline}), (\ref{mirrordiagonal}). It is thus sufficient to calculate the form of (\ref{sigSQ}) along the $x$-direction which in the $H$ plane corresponds to the imaginary axis $H=iJ$. The corresponding dependence in the circle along the $u$ direction follows from (\ref{hpsquaresig}), (\ref{hpcircle}) and reads
\begin{eqnarray} \label{circlesquaresig} 
\langle \sigma (u, u) \rangle &=& - \Bigl( {4 \over 1-u^2} \Bigr)^{1/8} \, {2 u^2 \over 1+u^4} \, .
\end{eqnarray}
Since $dw/dz = \sqrt{1+w^4} / \Lambda$ one finds for $|x| \ll {\cal W}$ where $u \ll 1$ the near center behavior
\begin{eqnarray} \label{squaresigx} 
\langle \sigma (x, x) \rangle_{\rm SQ} \to - \Lambda^{-1/8} \, 4^{1/8} \, 2 u^2 \to - \Bigl({4 \over \Lambda} \Bigr)^{1/8} \, 2 \Bigl( {x \over \Lambda} \Bigr)^2
\end{eqnarray}
along the $x$ direction and, due to symmetry, one finds the result (\ref{sigSQ}). This allows us to evaluate in Sec. \ref{OPEsq} the behavior of the response function near the center of the square.
 

\newpage
\begin{figure}
	\begin{center}
		\includegraphics[width=0.45\textwidth]{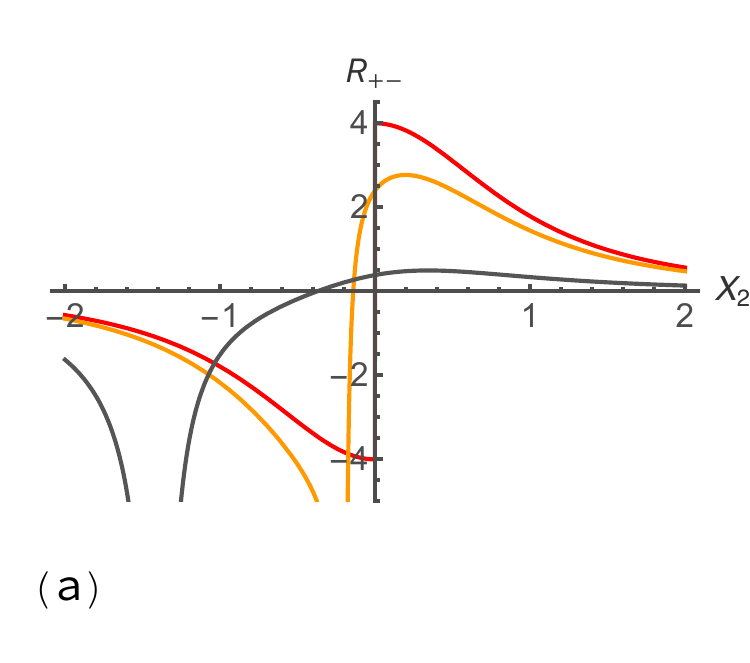}\\[1.0cm]
		\includegraphics[width=0.45\textwidth]{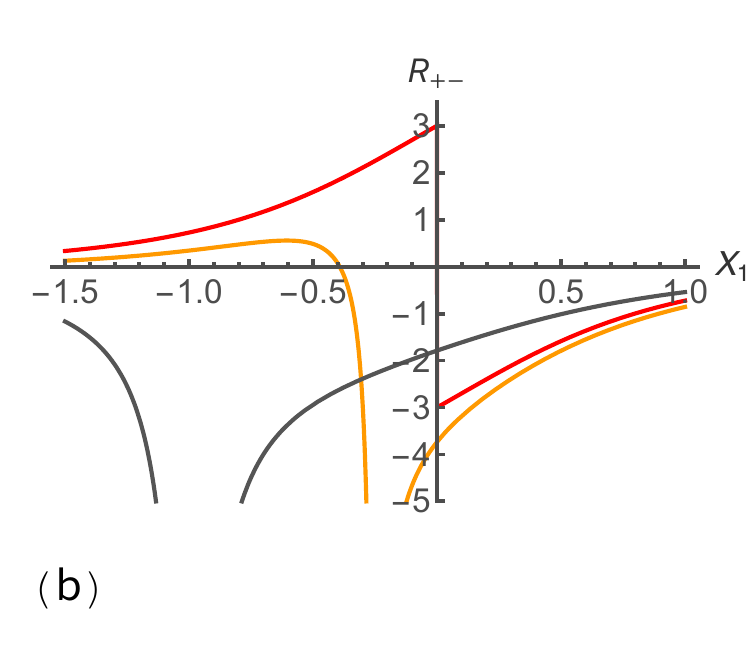}\qquad
		\includegraphics[width=0.45\textwidth]{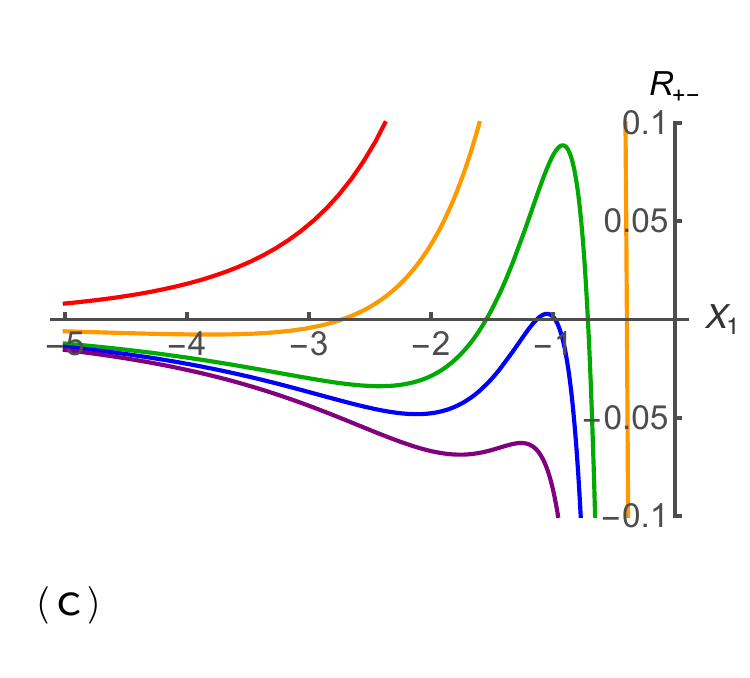}\\
	\end{center}
	\caption{Dimensionless response function $R_{+-} \equiv R_{+-}(X_1,X_2)$ along a line parallel to the $x$ axis in the upper half $x,y$ plane with $+-$ boundary condition as given in Eq. (\ref{se+-cumy0}). The red, orange, and grey curves in panel (a) show the $X_2$ dependence for $X_1$ fixed at $0, \, -0.2, \, -1.4$  while in panel (b) they show the $X_1$ dependence for $X_2$ fixed at $0, \, -0.25, \, -1$. For fixing $X_1$ and $X_2$ at 0 the upward and downward jumps, $4 \, {\rm sign} X_2$ and $-3 \, {\rm sign X_1}$ appearing in panels (a) and (b), respectively, are consistent with Eqs. (\ref{x1zero}) and (\ref{x2zero}) since $|\partial_x \, \langle \sigma (x, y_0) \rangle_{+-}|_{x=0} =   2^{1/8} / y_0^{(1/8)+1} $, see Eq. (\ref{sig+-}). On decreasing the fixed positions of $X_1$ and $X_2$ in panels (a) and (b), respectively, the corresponding $X_2$ and $X_1$ dependencies tend towards the $|X_2 -X_1|$ dependence for a uniform + boundary given in Eq. (\ref{+epssigprime}). In particular, the ``disordering {\it enhances} order'' regions adressed in paragraph (iii) of the Introduction that appear in (b) as $R_{+-}>0$ for $X_1 < 0$, decrease and vanish as the fixed locations $X_2$ of $\epsilon$ decrease from 0 via -0.25 to -1, so that $\langle \sigma \epsilon \rangle^{(\rm cum)}$ becomes negative for all $X_1 < 0$. The complexity of this process is displayed in more detail in panel (c) which shows the $X_1$-dependences for $X_2$ fixed at $0, \, -0.25, \, -0.36, \, -0.39$, and $-0.42$ in red, orange, green, blue, and purple: All curves with $X_2$ fixed must approach zero as $X_1 \to - \infty$. While for $X_2 =0$ (red curve) the approach is from above, see the remark below (\ref{half+-onepoint}), for all $X_2 < 0$ the approach is from below. }\label{FIG.1}
\end{figure}

\newpage
\begin{figure}
	\begin{center}
		\includegraphics[width=0.45\textwidth]{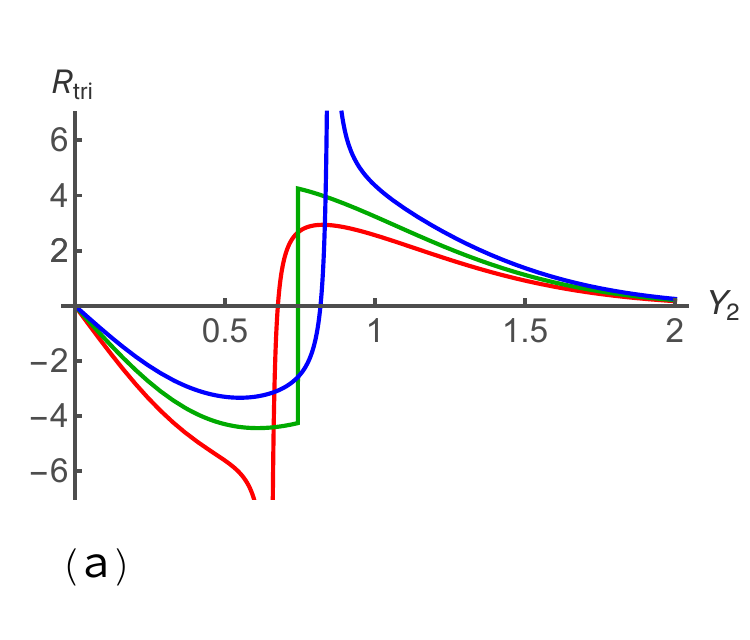}\qquad
		\includegraphics[width=0.45\textwidth]{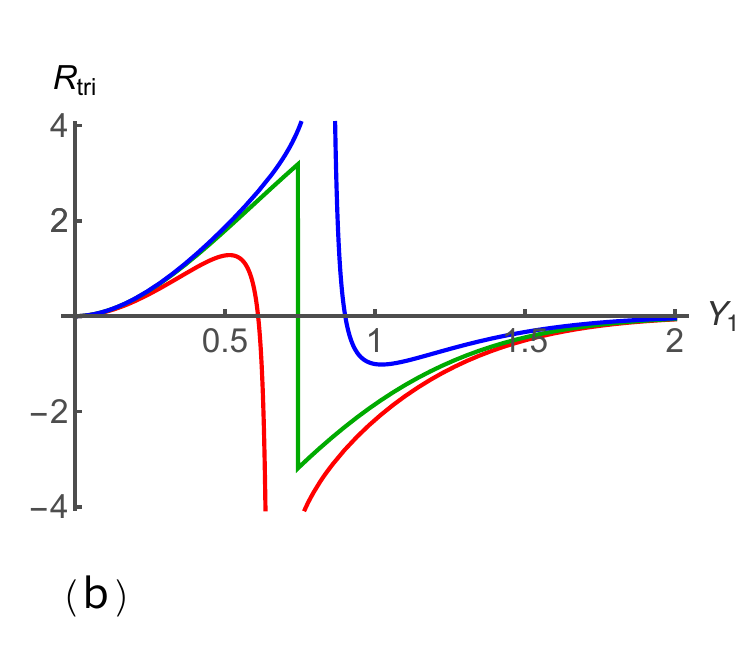}\\
	\end{center}
	\caption{The dimensionless response function $R_{\rm tri}$ on the vertical midline of the triangle defined in Eq. (\ref{tildeR}) and evaluated by means of (\ref{sigepstrafoprime}). The red, green, and blue curves in panel (a) show the dependencies of $R_{\rm tri}$ on $Y_2$ for $Y_1$ fixed at $0.65$, at $Y_0 = 0.743$, and at $0.85$. The response $R_{\rm tri}$ of the disorder to the up ordering at $Y_1 = Y_0$ (green curve) is negative and positive at $Y_2 < Y_0$ and  $Y_2 > Y_0$ since there the original order is in the up and down direction, respectively. For $Y_2$ close to $Y_0$ the $Y_2$-dependence reflects the asymptotic behavior given in Eq. (\ref{shorttri1at0}) with an upward discontinuity where $R_{\rm tri}$ jumps from $-(2 S_0)^{9/8} \equiv -(3^{3/4} 2^{2/3})^{9/8} = - 4.249$ to $(2 S_0)^{9/8}$. Unlike panel (a) in FIG 1 the $Y_2$-dependence is not antisymmetric about the discontinuity, and near the base line $Y=0$ and the corner $Y=Y_{\rm C} \equiv 3.196$ of the triangle it attains the behavior determined by Eqs. (\ref{y2to0}) and (\ref{y2toyC}). This decreases linearly from zero and approaches zero with the fifth power in the distance from the corner, respectively. For $Y_1 \not= Y_0$ (red and blue curves in panel (a)) the short distance singularity is $\propto |Y_2 - Y_1|^{-1} \,{\rm sign} (Y_1 -Y_0)$ and the prefactors of the near baseline and corner behaviors depend on $Y_1$ according to Eqs. (\ref{y2to0}) and (\ref{y2toyC}). Panel (b) shows the $Y_1$-dependence of the up order induced by the disorder imposed at $Y_2$ for $Y_2$ fixed at 0.65 (red), at $Y_0$ (green), and at 0.85 (blue). For $Y_2 = Y_0$ there is a downward discontinuity in the $Y_1$-dependence and the ratio of the discontinuities in panels (a) and (b) has the universal value of -4/3. Both for $Y_2 =Y_0$ and $Y_2 \not= Y_0$ the power law behaviors of $Y_1$ near the base line and the corner have exponents $2-(1/8)$ and $6-(1/8)$, respectively, with $Y_2$-dependent amplitudes according to Eqs. (\ref{y1to0}) and (\ref{y1toyC}). Note the $Y_1 < Y_0$ regions with ${\rm R}_{\rm tri} >0$ and the  $Y_1 > Y_0$ regions with ${\rm R}_{\rm tri} < 0$ where the magnitude of the order is {\it enhanced} by the disordering.} \label{FIG.2}
\end{figure}


\begin{thebibliography}{99}
\bibitem{ferro}Microscopically we have an Ising lattice model in mind with a ferromagnetic interaction between nearest neighbors.
\bibitem{spinreverse}Like the operator product $\sigma \times \epsilon$ on the left hand side of the OPE 
(\ref{next}) its right hand side must change sign on reversing all Ising spins and is composed of $\sigma$ and its descendents. Unlike the OPEs in (\ref{selfOPE}) there is no operator-free term in (\ref{next}).
\bibitem{singularope}The second term in the cumulant (\ref{respfunction}) is regular for $|{\bf r}_{12}| \to 0$ and does not contribute to the singular behavior of the terms given by the BOE (\ref{next}).
\bibitem{mixup}While $y_0$ denotes in Secs. \ref{intro}-\ref{hp+-} a given distance from the real axis of arbitrary value, in Sec. \ref{equitri} it denotes the value of $y$ on the midline of the triangle where $\langle \sigma \rangle_{\rm triangle}$ vanishes.
\bibitem{BOE}For the primary operators $\phi = \sigma$ or $\phi = \epsilon$ approaching a boundary section on the $x$-axis with boundary condition $a = +$ or $-$ the ``boundary operator expansion'' BOE has the form $$\phi(x, y)-\langle \phi \rangle_a \to \mu_a^{(\phi)}\, y^{2-x_{\phi}} T(x)  \, , \quad  y \to 0 \, .$$ Here the leading boundary operator $T$ is the stress tensor, $x_{\phi} = 1/8$ or $1$ are the scaling dimensions of $\sigma$ or $\epsilon$, and $$\mu_+^{(\sigma)} = - \mu_-^{(\sigma)} = -2^{1/8} \, , \,  \mu_+^{(\epsilon)} =  \mu_-^{(\epsilon)} = 4 .$$ For details see Eqs. (3.1)-(3.5) in Ref. \cite{BE21}.
\bibitem{complex}In order to facilitate the conformal transformations to other geometries, the exact results of Ref. \cite{BX} are written in complex notation.
\bibitem{COE}Consider the cumulant $\langle \phi \Phi\rangle_{\rm G}^{(\rm cum)}$ between two primary operators $\phi$ and $\Phi$ embedded in a geometry G with a corner, such as the triangle discussed in section \ref{equitri}. When $\phi$ approaches the tip of the corner and $\Phi$ is at a fixed finite distance from it, the cumulant takes the form $F \times \langle {\cal T} \Phi\rangle_{\rm G}^{(\rm cum)}$ where ${\cal T}$ is a ``corner operator'' which is located right at the tip of the corner and $F$ depends on the ``small distance'' $d$ of $\phi$ from the corner. If the two sides of the corner belong to the same universality class the distance-dependence is $F \propto d^{(2 \pi / \gamma)-x_\phi}$ where $\gamma$ is the angle enclosed by the corner, cp. Eq.(4.5) in Ref. \cite{E23}. In case of $\epsilon$ and $\sigma$ approaching the corner at point C of the equilateral triangle with $\gamma = \pi /3$ one finds $F \propto d^5$ and $d^{6-(1/8)}$, respectively, since $x_\phi = x_\epsilon =1$ an $x_{\sigma} =1/8$. Approaching a straight-line boundary such as the base side of the triangle can be viewed as approaching a corner with $\gamma = \pi$ which for $x_\phi = 1$ or $1/8$ yields the BOE-dependencies $d^1$ or $d^{2-(1/8)}$. Eqs. (\ref{y2toyC}) ff. and \red{ panels (a) and (b) in FIG. 2} for the triangle are in agreement with these dependencies.
\bibitem{center}The small value $y_0 = 0.2325 \times y_{\rm C}$ of the zero of $\langle \sigma \rangle_{\rm triangle}$ on the midline is not surprising. Only the base line has boundary condition + while the other two sides of the triangle both have $-$. So $y_0$ is definitely smaller than the value $y=y_{\rm center} =y_{\rm C}/3$ of the triangle's center where $\langle \sigma \rangle_{\rm triangle}$ obviously is negative. Note that the image in the $-+-$ half plane of $i y_{\rm center}$ is $h_{\rm center} = i \sqrt{3}$.
\bibitem{inverse}Once in a given geometry ${\cal G}$ the average $\langle T(z) \, \sigma (z_1, \, \bar{z}_1) \rangle_{\cal G}$ is known, the form of the corresponding average $\langle L_{-p} \, \sigma (z_1, \, \bar{z}_1) \rangle_{\cal G}$ follows most easily from averaging the inverse relation $$ T(z) \, \sigma (z_1, \, \bar{z}_1) = \sum\limits_{k=- \infty}^{\infty} \, (z-z_1)^{k-2} \, \bigl( L_{-k} \, \sigma (z_1, \, \bar{z}_1) \bigr)  $$
\bibitem{im-+-}For $h=ij$ the Moebius transformation (\ref{Möb}) reduces to $z = (j^2 -1 + 2ij)/(j^2 + 1)$ so that $z$ varies from $-1$ to 1 along the upper unit circle of the $+-$ plane as $j$ varies from 0 to $\infty$.
\bibitem{L1-+-}For the ${\cal a}$-term one needs $\langle L_{-1}\sigma \rangle_{-+-}$ for which Eq. (\ref{si-+-}) yields $\partial_{h_1} \langle \sigma \rangle_{-+-} = 2^{(1/8)-1} \times i$.
\end{thebibliography}
\end{document}